\documentclass[a4paper,
 reprint%
]{article}

\usepackage{amssymb,amsmath}
\DeclareMathOperator*{\argmax}{argmax}
\DeclareMathOperator*{\argmin}{argmin}

\usepackage{graphicx}
\usepackage{dcolumn}
\usepackage{bm}
\newcommand*{\plimsoll}{{\ensuremath{-\kern-4pt{\ominus}\kern-4pt-}}}
\usepackage{color}

\usepackage{adjustbox}
\usepackage{mathrsfs}
\newcommand{\setzeroone}{\{0,1\}}

\newcommand{\beginsupplement}{%
        \setcounter{table}{0}
        \setcounter{section}{0}
        \renewcommand{\thetable}{S\arabic{table}}%
        \renewcommand{\thesection}{S\arabic{section}}%
        \setcounter{figure}{0}
        \renewcommand{\thefigure}{S\arabic{figure}}%
     }

\usepackage{rotating}
\usepackage{graphicx}
\usepackage{dcolumn}
\usepackage{bm}
\usepackage{cancel}
\usepackage{graphics}

\usepackage{mathrsfs}

\usepackage{authblk}
\title{Evolution with recombination as a Metropolis-Hastings sampling procedure}
\author[1]{Jenny M. Poulton}
\author[2]{Lee Altenberg}
\author[3]{Chris Watkins}

\affil[1]{Foundation for Fundamental Research on Matter (FOM) Institute for Atomic and Molecular Physics (AMOLF), Amsterdam, 1098 XE, The Netherlands}
\affil[2]{Department of Mathematics, University of Hawai`i at M\=anoa, 2565 McCarthy Mall (Keller Hall 401A) Honolulu, HI 96822, United States}
\affil[3]{Department of Computer Science, Royal Holloway, University of London, Egham, Surrey, TW20 0EX, United Kingdom}
\affil[3]{Corresponding author; cjchwatkins@gmail.com}
\date{\today}
\begin{document}
\maketitle

\begin{abstract}
This work presents a population genetic
model of evolution, which includes haploid selection, mutation, recombination, and drift. The mutation-selection equilibrium can be expressed exactly in closed form for arbitrary fitness functions without resorting to diffusion approximations. Tractability is achieved by generating new offspring using n-parent rather than 2-parent recombination. While this enforces linkage equilibrium among offspring, it allows analysis of the whole population under linkage disequilibrium. 
We derive a general and exact relationship between fitness fluctuations and response to selection. Our assumptions allow analytical calculation of the stationary distribution of the model for a variety of non-trivial fitness functions. These results allow us to speak to genetic architecture, i.e., what stationary distributions result from different fitness functions. This paper presents methods for exactly deriving stationary states for finite and infinite populations. This method can be applied to many fitness functions, and we give exact calculations for four of these. These results allow us to investigate metastability, tradeoffs between fitness functions, and even consider error-correcting codes.

\end{abstract}

\maketitle


\newcommand{\bebi}{\textsf{BeBi}} 

\newcommand{\pop}{\mathbf{X}}  
\newcommand{\popprime}{\mathbf{X'}}
\newcommand{\alleles}{\{0,1\}}
\newcommand{\allpops}{\alleles^{N\times L}}
\newcommand{\pops}[2]{\pop_{#1}^{#2}}
\newcommand{\genome}{\mathbf{g}}
\newcommand{\locus}{\mathbf{c}} 
\newcommand{\Pu}{P_b} 
\newcommand{\Ps}{P_s} 

\newcommand{\fitnessi}{f^i}
\newcommand{\phii}{\phi^i}

\newcommand{\numones}{\mathbf{n_1}}

\newcommand{\perfect}{\textsf{perfect}}
\newcommand{\Zperfect}{Z^{\textsf{  \perfect}}}
\newcommand{\Pperfect}{P^{\textsf{  \perfect}}_u}
\newcommand{\Psperfect}{P^{\textsf{  \perfect}}_s}

\newcommand{\prefix}{\textsf{prefix}}  
\newcommand{\Zprefix}{Z^{\prefix}}
\newcommand{\Pprefix}{P^{\prefix}_u}
\newcommand{\Psprefix}{P^{\prefix}_s}

\newcommand{\onemax}{\textsf{onemax}}
\newcommand{\Zonemax}{Z^\onemax} 
\newcommand{\Ponemax}{P^\onemax_u}
\newcommand{\Psonemax}{P^\onemax_s} 

\newcommand{\sfsum}{\textsf{sum}}
\newcommand{\Zsum}{Z^{\textsf{sum}}} 
\newcommand{\Psum}{P^{\textsf{sum}}}
\newcommand{\Pssum}{P^{\textsf{sum}}_s} 

\newcommand{\fperfect}{f^{\perfect}}
\newcommand{\fprefix}{f^{\prefix}}
\newcommand{\foneerror}{f^{\oneerror}}
\newcommand{\fsum}{f^{\sfsum}}

\newcommand{\oneerror}{\textsf{oneerror}}

\newcommand{\Poneerror}{P^\oneerror_u}
\newcommand{\Psoneerror}{P^\oneerror_s}
\newcommand{\Zoneerror}{Z^{\textsf{\oneerror}}}

\newcommand{\fraconesatlocus}{\theta} 

\newcommand{\bx}{\mathbf{x}} 
\newcommand{\by}{\mathbf{y}}
\newcommand{\bg}{\mathbf{g}} 

\newcommand{\vectheta}{\vec{\theta}}
\newcommand{\vecalpha}{\vec{\alpha}}
\newcommand{\vecu}{\vec{u}}

\newcommand{\maximiser}{\Omega}

{\bf Moran process, Dirichlet process, Detailed balance, Genetic Architecture}

{\bf Declarations of interest: none.}

\section{Introduction}

Mathematical models of natural selection, mutation, recombination, and drift are notorious for becoming intractable when scaled up beyond the simplest situations. Intractability is usually dealt with by making various simplifying assumptions. The art of these assumptions is to afford tractability without abrogating the biological relevance of the model. 

Population genetic models of finite populations typically generate population dynamics as Markov chains, which can be potentially analyzed with the spectral theory of finite matrices. However, these models have large state spaces and complicated matrix elements, so approximation and other methods are necessary. These approaches are used at the outset in the finite population Wright-Fisher model \cite{wright1931evolution} and Moran model \cite{moran1958random}. For example, \cite{baake2003exactly} develops a Moran model incorporating mutation and recombination, which is exactly solvable only for restricted types of selection.

Here we investigate a novel direction for simplifying assumptions that allow us to obtain analytical results for multiple loci under recombination, selection, mutation, and drift in a Moran-based model introduced in \cite{lember2020evolutionary,watkins2014}. Our simplifications are straightforward. The main simplification is in modeling sexual breeding through recombination. Unlike existing models, which use two-parent recombination, our model has n-parent recombination. Genetic material from the whole population is combined when breeding a new offspring. We suggest that this is a  natural abstraction of evolution with sexual reproduction. This abstraction does preserve one essential property of recombination: genetic elements are drawn from parents independently from one another. 

A major obstacle to tractability in models that include recombination is the persistence of linkage disequilibrium in the dynamics. Linkage disequilibria is where alleles at different loci are associated within genomes. In numerous models of unlinked or loosely linked loci, linkage equilibrium is assumed to make the analysis tractable \cite{turelli1990dynamics,carter2005role,malaspinas2011match,novak2017does}. Our analysis can accommodate linkage disequilibrium in the entire population. However, our simplification for tractability is that new offspring are created in linkage equilibrium. Alleles for each locus are drawn from the entire population, independently from the draw at the other loci. Organisms are haploid, and their single genome determines fitness.

This sampling assumption, which is effectively $n$-parent recombination, and several other modeling assumptions, produce an implementable evolutionary model that exactly satisfies detailed balance at its stationary distributions. The stationary distribution of this evolutionary model can be given in closed form for any fitness function. This work shows how to calculate the stationary distribution for both finite and infinite populations. It is notable that despite being Markov chains, previous models for long multi-locus genomes with recombination and arbitrary selection do not exhibit reversibility, and therefore neither do their evolutionary stationary distributions satisfy detailed balance.

Many people have remarked on the similarity between the replicator equation, and Bayesian updating \cite{harper2009replicator,watson2016can,burger2000mathematical}.  
Our evolutionary model is not just reminiscent of Bayesian updating --- it is precisely a Monte-Carlo procedure for sampling from the posterior distribution of a well-known Bayesian model, a factorial Dirichlet process. The interpretation of our model is entirely non-Bayesian, but the formal identity to established Bayesian modeling is exact. The stationary distribution of populations is a product of two factors: a `breeding' factor, analogous to prior probability, which depends only on the mutation rate and the population size; and a `selection' factor, which depends only on the definition of fitness used, and which is analogous to a likelihood. 

A fundamental property of any evolutionary model satisfying detailed balance is that, at mutation-selection equilibrium, it is impossible to determine the direction of time from the sequence of overlapping populations. On a microscopic level, biological evolution is not reversible because the time direction of genealogies can be determined: it is possible to identify parent-child-parent triples so that it is possible to distinguish between the genomes of children and their parents. In this way, one may identify the direction of time. Our analysis does not keep track of genealogies, only population compositions, for which the dynamics are reversible.

Instead of questions of population history and lineage, we are interested in genetic architecture. For example, what stationary distributions of genomes are induced by different classes of fitness functions?
The relationship of organismal fitness to genotype 
is complex: what are the evolutionary effects of different mappings from genotype to fitness? Our pan-population recombination model is needed to investigate these questions because the stationary distribution is hard or impossible to calculate for models with a more realistic breeding model. We explore this with calculations for some example fitness functions in section \ref{sec:fourfitnessfunctions} below. 

\medskip

In this paper, we make the following specific contributions. 
First, in section \ref{chainalgorithm}, we introduce a continuous-time Moran process, with recombination, mutation, and selection, implementable as an individual agent-based simulation, 
which exactly satisfies detailed balance.

Second, we show that our model is not just analogous to a Bayesian model - it \emph{actually is} a Bayesian model of a well-known type. Section \ref{sec:Bayesian} describes how our evolutionary model is formally identical to a variant of a well-known Bayesian model, the Dirichlet process mixture model \cite {neal2000markov, teh2010dirichlet}, with a prior that is a product of independent Dirichlet processes, a simpler version of the processes considered in \cite{griffiths2011indian}.    However, the motivation and interpretation of our model is entirely different from these Bayesian models. We are not interested in data analysis but in an analytically tractable abstract model of evolution. Our model may give insight into adaptation with complex fitness functions.  

Third, in section \ref{sec:fitnessfluctuations},  we establish a universal relationship between the fluctuations of population fitness at stationarity and the gradient of a population's mean fitness with respect to the selection intensity. This relationship is valid for all fitness functions. As far as we are aware, this analysis of fitness fluctuations is new. 

Fourth, in section \ref{sec:fourfitnessfunctions}, we explicitly show how to exactly calculate the stationary distribution for four non-trivial fitness functions for both finite and infinite populations. These calculations provide insight into cases with meta-stable regimes, where accurate simulations require long runs. 

Finally, in section \ref{sec:errorcorrecting}, we demonstrate that mutational robustness in the form of 
error-correcting codes, in particular, Hamming codes, can decrease the genetic load or, equivalently, allow a given fitness level to be maintained in the face of mutation with less intense selection.

\section{The evolutionary model}
\label{OurModel}

We consider a population \(\pop\) of \(N\) haploid genomes with \(L\) biallelic loci, each represented as a binary vector of length \(L\). There are therefore \(2^{N\times L}\) possible population states. We consider the population of $N$ genomes as a $N\times L$ matrix \(\pop\) of binary values. Each row represents a genome, and each column represents the values at a particular position or locus across all genomes. 

We denote genome $i$ as  \(\bg_{i}=(\pop_{i1},\pop_{i2},...,\pop_{iL})\), the $i$'th row of $\pop$. The $j$'th column of $\pop$, which represents all values at the $j$'th locus in each of the \(N\) genomes, will be denoted by  \(\locus_{j}=(\pop_{1j},\pop_{2j},...,\pop_{Nj})\).  
The $k$'th element of locus \(j\) will be denoted $\locus_j(k)$, and the $m$'th element of genome \(i\) will be denoted $\bg_i(m)$.

We consider evolution as a continuous time Markov chain. At each state transition, one genome is selected to die, and a new genome is immediately `bred' from the remaining genomes to replace it. The generations are overlapping: only one genome changes at each jump in the chain.

\subsection{Fitness and lifetime}
In our model, the fitness of a genome is modeled as its expected lifetime in the Markov process. We write the fitness of a genome \(\bg\) as \(f(\bg)\), where $f$ is a `fitness function' that maps binary vectors of length $L$ to strictly positive real numbers. We will consider different choices of $f$ below. 

It will frequently be convenient to parametrise $f$ in terms of a `weight' function $\phi$ and a parameter $\beta \ge 0$, such that $f(\bg) \equiv \exp(-\beta\phi(\bg))$. For a given $\phi$, varying $\beta$ corresponds to varying selection intensity. 

 When $\beta=0$, all genomes have expected lifetime 1, there is no selection effect because all genomes have the same fitness. For small $\beta>0$, genomes differ only slightly in fitness, and selection is weak. For large $\beta$, some genomes may have long lifetimes, while others `die' almost immediately: increased values of $\beta$ induce increased differences in genome lifetimes for given fitness weights $\phi$. 

A genome $\bg$ with weight $\phi(\bg)$ has expected lifetime $\exp(-\beta \phi(\bg)))$. In the Markov process (section \ref{chainalgorithm}) below, the lifetime of $\bg_i$ is an exponentially distributed random variable, with mean $\exp(-\beta \phi(\bg_i)))$. We assume that a genome's fitness is independent of other genomes in the population and remains constant over its lifetime. The lifetime for each individual with genome $\bg_i$ is set at birth by drawing a value from its exponential lifetime distribution $\phi(\bg_i)$.

The probability that the living genome \(\bg_i\) dies in a time interval $dt$ is $\beta \phi(\bg_i) dt$. It follows that, in a population of living genomes $\bg_1, \ldots, \bg_N$,
\begin{equation}
    P({\textsf{next genome to die is }} \bg_i)=\frac{\frac{1}{f(\bg_i)}}{\frac{1}{f(\bg_1)}+...\frac{1}{f(\bg_i)}+...\frac{1}{f(\bg_{N})}}.
\end{equation}
It follows that the expected time until the next transition in the Markov chain of populations is 
\begin{equation}
   \frac{1}{\exp(\beta\phi(\bg_1)) + \cdots + \exp(\beta\phi(\bg_N))}.
\end{equation}

Adding a constant to $\phi$, equivalent to multiplying all fitnesses by a constant factor, does not affect the stationary distribution. 

Since $f(\bg) = \exp( -\beta \phi(\bg))$, $f$ is invariant if $\phi$ is multiplied by some constant $c$, and $\beta$ is divided by the same constant. 

\subsection{Breeding as Gibbs sampling}\label{Urn section}

Our model of breeding is analogous to recombination through sexual reproduction. However, instead of breeding from two parent genomes, a new genome is bred by combining randomly chosen elements from \emph{all} genomes in the population. We make this modeling choice for mathematical simplicity: with this choice, the Markov chain of populations satisfies detailed balance, and the stationary distribution factorizes into a simple form. 

Once genome $\bg_i$ dies, a new genome is instantly bred to replace it. Fitness does not affect breeding. The new genome $\bg^\prime_i$ is resampled from the remaining existing genomes $\bg_1, \ldots, \bg_{i-1}, \bg_{i+1}, \ldots, \bg_N$ without any dependence on their fitnesses. 

\subsubsection{Mutation}
\label{sec:Mutation}

In this paper, we parameterize mutation in two equivalent ways: using mutation rates and using concentration parameters of a Dirichlet process. Parametrizing mutation as a Dirichlet process allows us to directly connect with nonparametric Bayesian statistics and derive the mutation-selection equilibrium in closed form. 

A conventional notation to define mutation rates is as a column-stochastic allele transition matrix $\mathbf{T}$, defined as: 
\begin{align}
\mathbf{T} = (1-\mu) \mathbf{I}  + \mu \mathbf{M} \\
\mathbf{M} = \begin{pmatrix}
1-u_{10} & u_{01} \\
u_{10} & 1-u_{01}
\end{pmatrix}
\end{align}
where $T_{ij}$ is the probability of a transition from allele $j$ to allele $i$, and  $u_{01}$ is the probability that a 1-allele mutates to a 0-allele, given that a mutation occurs. 

Let us define: 
\begin{align} 
u_0 = \mu u_{01}\\
u_1 = \mu u_{10}\\
u = u_0 + u_1 
\end{align}
We may now re-express $\mathbf{T}$ as: 
\begin{equation}
    \mathbf{T} = 
    \begin{pmatrix}
    (1 - u) + u_0 & u_0 \\
    u_1 &  (1 - u) + u_1 
    \end{pmatrix}
\end{equation}
We can describe the action of $\mathbf{T}$ in two ways. First, as mutation: a new allele is `bred' by copying an existing allele from a parent, and then $\mathbf{T}$ is applied to stochastically mutate the allele. Second, as sampling from a base distribution: when a new allele is bred, it is copied from some parent in the population with probability $1-u$, and the new allele instead is sampled from a `base distribution' with probability $u$, with $p(0) = \frac{u_0}{u}$, and $p(1) = \frac{u_1}{u}$. The sampling from the base distribution does not depend on the current composition of the population. Therefore, in our model, mutation can be considered as sampling from a fixed distribution of possible alleles instead of transforming one allele into another. 

In this paper, we only consider the case of two alleles (0 and 1). In this case alone, the two views of mutation are mathematically equivalent. For more than two possible alleles, the second `sampling' view of mutation imposes restrictions on the structure of the mutation matrix: this constraint is discussed in \cite{lember2020evolutionary}, but we do not consider it here. 

Finally, we reparametrize $u_0$ and $u_1$ as concentration parameters: 

\begin{align}
\begin{split}    
\alpha_0 &:= (N-1)\frac{u_0}{1-u}
\end{split}
\begin{split}
    \alpha_1 &:= (N-1)\frac{u_1}{1-u}
\end{split}  
\end{align}

\begin{equation}
    \alpha := \alpha_0 + \alpha_1
\end{equation}
\noindent Note that $\alpha$ is close to the standard genetic parameter $Nu$. To convert from $\alpha$ back to $u$ we have: 

\begin{align}
\begin{split}
u_0& := \frac{\alpha_0}{\alpha + N - 1}
\end{split}
\begin{split}
u_1& := \frac{\alpha_1}{\alpha + N - 1}
\end{split}
\end{align}

The point of introducing $\alpha, \alpha_0, \alpha_1$ is to enable the stationary distribution of the number of 0 and 1 alleles to be written in a convenient closed form, which we now do. 
\subsubsection{Breeding a new allele}

Consider a population of $N$ genomes, each consisting of one allele, which may be either 0 or 1. We repeatedly randomly remove a single allele -- it `dies' -- and a new individual is bred. This breeding happens by sampling the new allele conditionally on the $N-1$ existing alleles in the following way. With probability $\frac{\alpha}{\alpha+N-1}$ there is a mutation, and a new allele is sampled from the base distribution $p(0) = \frac{\alpha_0}{\alpha}$, $p(1) = \frac{\alpha_1}{\alpha}$. With probability $\frac{N-1}{\alpha + N - 1}$, the new allele is a copy of an allele uniformly selected from the $N-1$ existing alleles. With our definitions of $\alpha_0$ and $\alpha_1$, this is exactly breeding with mutation rates $u_0$ and $u_1$.

If individuals are chosen uniformly at random to die, what is the stationary distribution of the numbers of 0 and 1 alleles?

When sampling a new allele given a population of $n$ existing alleles, $x_1, \ldots, x_n$, comprising $n_0$ 0-alleles and $n_1$ 1-alleles, with $n_0 + n_1 = n$, with given $\alpha_0,\alpha_1$, we have : 
\begin{align}
    \begin{split}
    p(x_{n+1} = 0 \mid x_1, \ldots, x_n) = \frac{\alpha_0 + n_0}{\alpha + n},
    \end{split}
    \begin{split}
    p(x_{n+1} = 1 \mid x_1, \ldots, x_n) = \frac{\alpha_1 + n_1}{\alpha + n}.
    \end{split}    
\end{align} 

The key observation of \cite{blackwellmacqueen} is that, keeping $\alpha_0$ and $\alpha_1$ fixed and not varying with $n$, the sequence of alleles sampled by this process is exchangeable. It follows that the process of randomly deleting an allele from a population of $N$ and resampling (Gibbs sampling) must have the same distribution as that of a sequence of alleles sampled beginning with a population of size zero. The probability of a particular locus $\locus=x_1, \ldots, x_N$ containing $n_0$ 0-alleles and $n_1$ 1-alleles is readily seen to be
\begin{equation}
    \Pu( \locus) = \frac{\alpha_0(\alpha_0+1)\cdots(\alpha_0+n_0-1) \, \times \, \alpha_1(\alpha_1+1)\cdots(\alpha_1+n_1-1) }{\alpha( \alpha+1) \cdots (\alpha+N-1)}
\end{equation}
by rearrangement of terms. Although we consider sampling starting with an empty population to calculate the stationary probability of the locus, the mutation rate for each conditional sampling from $N-1$ individuals is $u= \frac{\alpha}{\alpha+N-1}$.

We define \(\Pu(n_{0},n_{1})\) as the probability of  \(n_{0}\) alleles of type \(0\)  \(n_{1}\) alleles of type one. The stationary distribution is the Beta-Binomial distribution 
\begin{equation}
     \Pu(n_{0},n_{1})={n \choose n_{0}} \frac{\alpha_{0}(\alpha_{0}+1)...(\alpha_{0}+n_{0}-1)\times\alpha_{1}(\alpha_{1}+1)...(\alpha_{1}+n_{1}-1)}{(\alpha_{0}+\alpha_{1})(\alpha_{0}+\alpha_{1}+1)...(\alpha_{0}+n_{0}+n_{1}-1)}.
\end{equation}   
By using the rising factorial function \((\alpha_{j})_{n}=\alpha_{j}(\alpha_{j}+1)(\alpha_{j}+2)...(\alpha_{j}+n-1)\), the probability may be written as:
\begin{equation}
\label{eq:betabinomial}
     \Pu(n_{0},n_{1})={n \choose n_{0}} \frac{(\alpha_{0})_{n_{0}}(\alpha_{1})_{n_{1}}}{(\alpha_{0}+\alpha_{1})_{n_{0}+n_{1}}}.
\end{equation}

\subsubsection{Breeding a multi-locus genome} 

During the evolutionary process, a whole genome is removed at once, and a new genome is bred from the remaining $N-1$ genomes to replace it. 

Our simplifying abstraction is that at each locus of the new genome, the new allele is sampled independently of the new values at other loci. It follows there is a separate Blackwell-MacQueen process at each locus, and these processes are independent so that in the absence of selection (that is if deaths occur at random), the breeding probability of a given population \(\pop\) is then
\begin{equation}
    \Pu(\pop)=\prod_{j=1}^{L}\Pu({\locus}_{j}).
\end{equation}
This distribution functions as a `prior' distribution, with fitness playing the role of likelihood. 

\subsection{A Continuous-Time Evolutionary Process}
\label{chainalgorithm}
We model evolution as a continuous-time Markov chain of overlapping populations, where successive populations differ by only one individual. An algorithm for simulating the Markov chain is as follows:

\begin{enumerate}
    \item Given: genomes $\bg_1, \ldots, \bg_N$ with fitnesses $f_1, \ldots, f_N$
    \item For $1 \le i \le N$ do: sample $\omega_i \sim \textsf{Exponential}(\frac{1}{f_i}) $\\
    \emph{// $\omega_i$ is the time of death of $\bg_i$}
    \item For $l = 1$ to $\infty$ do: \emph{// $l$ indexes transitions}
    \begin{enumerate}
        \item Let $i= \argmin(\omega_1,\ldots, \omega_N)$, and let $\tau_l = \omega_i$ \\
        \emph{// \(\bg_i\) is the first genome to die; time of $l$'th transition is $\tau_l =\omega_i$}
        \item Reject $\bg_i$ 
        \item Sample new $\bg_i \sim \Pu(\cdot\mid \bg_1, \ldots, \bg_{j-1}, \bg_{j+1}, \ldots, \bg_N)$
        \item $f_i := f(\bg_i)$ 
        \item Sample $\epsilon \sim  \textsf{Exponential}(\frac{1}{f_j})$ \emph{//compute lifespan of new $\bg_i$}
        \item Set $\omega_i := \tau_l + \epsilon$ \\
        //sets death date of new $\bg_i$ as current time $\tau_l$ plus lifespan $\epsilon$
    \end{enumerate}
\end{enumerate}

\subsection{Stationary distribution of the Markov chain of populations}

 Since each population depends only on the previous population, the sequence of populations is a Markov chain. This chain has a finite state space because we consider states consisting of $N$ genomes of length $L$. Moreover, the chain is irreducible because mutations can enable any population to transition into any other population over a sequence of $N$ transitions. It follows that the chain has a unique stationary distribution.

We make an ansatz that the stationary distribution $\Ps$ factorizes into $\Pu$ reweighted by the product of fitnesses of genomes in the population:
\begin{equation}
\label{eq:Pss} 
    \Ps(\pop)=\frac{1}{Z}\Pu(\pop)\prod_{i=1}^{n}f(\bg_i),
\end{equation}
where $Z$ is a normalizing factor that sums over all possible populations $\pop$:
\begin{equation}
    Z(N,L, f) = 
    \sum_{\pop \in \allpops} \Pu(\pop) \prod_{i=1}^{n}f(\bg_i)
\end{equation}
Note that $\Pu$ depends on $\alpha_0$, $\alpha_1$ and $Z$ depends upon $N$, $L$, $\alpha_0$, $\alpha_1$ and $f$, but we do not explicitly state these dependencies throughout.
We now show that this ansatz satisfies the detailed balance equations, which proves that it is indeed the stationary distribution.

Consider two populations $\pop=(\bg_1, \ldots, \bg_N)$ and $\pop^\prime=(\bg_1, \ldots, \bg_i^\prime, \ldots, \bg_N)$, which differ only in their $i$'th genome. Direct transitions are only possible between pairs of populations, such as $\pop$ and $\pop^\prime$, that differ in only one genome. Hence, we need only consider transitions between such populations to establish detailed balance. 

The detailed balance equations are, for each pair $\pop, \pop^\prime$ differing in at most one genome: 
\begin{equation}
\label{eq:detailedbalance}
\Ps(\pop) T(\pop \rightarrow \pop^\prime) = 
\Ps(\pop^\prime) T( \pop^\prime \rightarrow \pop)
\end{equation}
where $T(\pop \rightarrow \pop^\prime)$ is the rate for population $\pop$ to transition to $\pop^\prime$.

For a transition from $\pop$ to $\pop^\prime$ to occur, genome $\bg_i$ in $\pop$ must die, and then genome $\bg_i^\prime$ must be bred from the remaining genomes, which we denote as $\pop_{\backslash i}= \bg_1, \ldots, \bg_{i-1}, \bg_{i+1}, \ldots,\bg_N$. 

Let us write the fitnesses $f(\bg_1),\ldots,f(\bg_N)$ as $f_1,\ldots, f_N$  for brevity. The hazard rate for $\bg_i$ to die is then $\frac{1}{f_i}$. The probability of breeding $\bg_i^\prime$ from $\pop_{\backslash i}$ is $\Pu(\bg_i^\prime \mid \pop_{\backslash i})$, so that
\begin{equation}
T(\pop \rightarrow \pop^\prime) = \frac{1}{f(\bg_i)} \Pu( \bg_i^\prime \mid \pop_{\backslash i})
\end{equation}
hence we obtain
\begin{align}
    \Ps(\pop) T(\pop \rightarrow \pop^\prime) & = \frac{1}{Z}\Pu(\pop) f_1\cdots f_N \frac{1}{f_i}\Pu( \bg_i^\prime\mid \pop_{\backslash i}) \\
    & = \frac{1}{Z}\Pu(\pop)\Pu( \bg_i^\prime\mid \pop_{\backslash i}) f_1\cdots f_{i-1}f_{i+1}\cdots f_N\\
\intertext{and since $\Pu(\pop)=\Pu(\pop_{\backslash i})\Pu(\bg_i\mid\pop_{\backslash i})$,} 
    & = \frac{1}{Z}\Pu(\pop_{\backslash i}) \Pu(\bg_i\mid \pop_{\backslash i})\Pu( \bg_i^\prime \mid \pop_{\backslash i}) f_1\cdots f_{i-1}f_{i+1}\cdots f_N
\end{align}
The last expression is symmetric in $\bg_i$ and $\bg_i^\prime$, showing that detailed balance holds, which proves that the ansatz in equation \ref{ep:Pss} is indeed the stationary distribution.

The stationary distribution can be written as:
\begin{equation}
        \Ps(\pop)=\frac{1}{Z}\prod_{j=1}^{l}\Pu(\locus_{j})\prod_{i=1}^{n}f(\bg_i),
\end{equation}
which shows that the stationary probability of a population is proportional to the product of column-factors, one calculated for each column of $\pop$, $\locus_1, \ldots,\locus_L $, multiplied by the product of row-factors, one calculated for each row of $\pop$, $\bg_1, \ldots, \bg_N$. The column-factors relate only to breeding: they depend only on the population size and mutation rate. The row factors are individual genomes' fitnesses (or expected lifetimes). In this model, these fitnesses are assumed to be a function of the genome sequence only and not dependent on the other genomes in the population. 

While the breeding processes at each locus are independent, the fitness functions can cause dependencies between the loci on the genome at the point of death. Hence, dependencies between columns of $\pop$ are introduced by selection. In genetic language, breeding without selection gives a population in linkage equilibrium, but selection introduces dependencies between loci. 

\subsection{A factorial Dirichlet process model}
\label{sec:Bayesian}

Readers familiar with Bayesian nonparametrics may recognize that this stationary distribution occurs as the posterior distribution over latent variables in factorial Dirichlet process models, as in \cite{neal2000markov,teh2010dirichlet,griffiths2011indian}. 

Suppose we have $N$ data samples $y_1, \ldots, y_N$. For each sample, we wish to infer $L$ latent attributes. Each attribute may be either present or absent, coded as $1$ or  $0$, respectively. The latent variables form a $N$ by $L$ matrix $\pop$ of binary values, each row corresponding to a sample, each column corresponding to an attribute.
We may formulate a model as follows: 

Given: sample size $N$, number of factors $L$, Dirichlet parameters $\alpha_0, \alpha_1 > 0$, and a likelihood function $f$. 
\begin{align}
\theta_j & \sim \textsf{Beta}(\alpha_0, \alpha_1) ~ \text{for $1 \le j \le L$}\\
\pop_{ij} & \sim \textsf{Bernoulli}(\theta_j) ~ \text{for $1\le j \le L$, $1 \le i \le N$}\\
y_i & \sim p(\cdot \mid \pop_{i1}, \ldots,\pop_{iL}) ~ \text{for $1 \le i \le N$}
\end{align}
where we use the fact that the Beta-Binomial distribution is a mixture of binomial distributions: 
\begin{equation}
    \bebi(\locus) = \int_{\theta=0}^{1} P( \locus \mid \theta) \mathsf{Beta}(\theta; \alpha_0, \alpha_1) d\theta.
\end{equation}

Although we have constructed an evolutionary model with the form of a well-known Bayesian non-parametric model, our justification is in the reverse direction of every step of the standard Bayesian argument. In Bayesian statistics, one starts with data and prior beliefs about the data. These beliefs are expressed as a generative model: a prior distribution over latent parameters and a likelihood function that gives the probability of the data according to the model. The Bayesian wishes to know the distribution of the latent parameters, given the data. The full Bayesian analysis would require summing over all values of the latent parameters; as this is typically infeasible, it is then necessary to devise a Monte-Carlo Markov chain (MCMC) sampling scheme to sample sets of latent parameters. The Bayesian model above can be described by 
\begin{align}
p(\pop \mid \alpha_0, \alpha_1, \by) &= \frac{p(\pop \mid \alpha_0, \alpha_1)p(\by \mid \pop) }{p(\by \mid \alpha_0, \alpha_1)}\label{eq:bayesposteriorx1}\\
& = \frac{1}{p(\by \mid \alpha_0, \alpha_1)} \prod_{j=1}^L p(\locus_j \mid \alpha_0, \alpha_1) \prod_{i=1}^N p( y_i \mid \pop_i ).
\label{eq:bayesposteriorx2}
\end{align}
Equation \ref{eq:bayesposteriorx2} has a simple form: the posterior probability of $\pop$ is proportional to the product of the prior probabilities of its columns $\locus_j$ and the likelihoods of its rows $\bg_i$. However, because of the large number of possible discrete values of $\pop$, it is not straightforward to sample from its posterior distribution nor to compute the evidence $p(\by )$. Typically a Monte-Carlo method must be used. The continuous-time Markov chain described in section \ref{chainalgorithm} above is, by the standards of Monte-Carlo methods in statistics, a straightforward and naive method of sampling from the posterior over the latent parameters $\pop$.

Our reasoning is, in fact, the opposite of this. To develop our genetic model, we start instead with our specific Monte-Carlo Markov chain sampling algorithm, outlined above, which is abstractly similar to sexual reproduction with recombination, selection, and mutation. We ensure that the sampling algorithm satisfies detailed balance through careful design choices. Consequently, its stationary distribution is easy to derive and can be written in a simple closed form. The stationary distribution factorizes cleanly as a `breeding' term multiplied by a `selection' term. To our surprise, this stationary distribution is formally identical to the posterior over latent parameters of a well-known non-parametric Bayesian model. The prior distribution of the model corresponds to the probability of breeding a specific genome in our model, and the fitness function used for selection corresponds formally to the likelihood function of the model. Happily, we have started with an abstract evolutionary model and found that it can be viewed as an MCMC sampling procedure for latent parameters of a well-known non-parametric Bayesian model (a product of Dirichlet processes). 

We give $\pop$ a physical meaning: a set of \(N\) genomes of length \(L\). Each element $\pop_{ij}$ corresponds to locus $j$ in genome $i$, with each row of $\pop$ corresponding to one genome. We replace the likelihood with a fitness function. A genome's fitness is denoted by $f(\bg_i)$, with all genomes having the same fitness function \(f\). It is as if the data set consisted entirely of a single value so that $\by_1 = \cdots = \by_N$, and $f(\bg_i) = p(\by_i \mid \pop_i)$ for each $i$. Our fitnesses can have any value strictly greater than zero: in our model, $f(\bg_i)$ can be interpreted as the expected lifetime of the individual $\bg_i$. So, instead of the standard Bayesian equations \ref{eq:bayesposteriorx1},\ref{eq:bayesposteriorx2} above, we write instead: 
\begin{equation}
p(\pop \mid \alpha_0, \alpha_1) = \frac{ p(\pop \mid \alpha_0, \alpha_1)f(\pop)}{Z(f,\alpha_0,\alpha_1)}
= \frac{1}{Z(f,\alpha_0,\alpha_1)} \prod_{j=1}^L p(\locus_j \mid \alpha_0, \alpha_1) \prod_{i=1}^N f( \bg_i )
\label{eq:modelx2}
\end{equation}
where
\begin{equation}
Z(f,\alpha_0,\alpha_1) =  \sum_\pop p(\pop \mid \alpha_0, \alpha_1)f(\pop),
\end{equation}
is a normalizing constant. This sampling method seems as plausible a model of evolution as many other proposed simple algorithms such as \cite{harvey2009microbial}, but with the advantage that the Markov chain is reversible. 

\section{Fitness fluctuations} 
\label{sec:fitnessfluctuations}

With finite population size, how large are the fluctuations in population fitness at stationarity? This question is interesting because if the fitness of a finite population of organisms in nature fluctuates too much, extinction is possible. We derive a general relationship between fitness fluctuations and sensitivity of mean log fitness to  $\beta$.

Recall that we may write fitness in terms of either $f$ or $\phi$, with $f(\bg) = \exp( -\beta \phi(\bg))$, where $\beta>0$ specifies the strength of selection. 
\begin{align} 
\Ps( \pop )& =  \frac{1}{Z} \Pu(\pop) \exp\left(-\beta \sum_{i=1}^N \phi(\bg) \right) \nonumber\\
\intertext{Using the notation $\phi(\pop):=\sum_{i=1}^N \phi(\bg)$ we write}
&= \frac{1}{Z} \Pu(\pop) \exp(-\beta\phi(\pop))
\end{align}
where the normalizing constant $Z$ is
\begin{align}
Z(N,L,\vecalpha,\beta,\phi) & = \sum_{\pop\in\alleles^{N\times L}}\Pu(\pop) \exp(-\beta\phi(\pop))
\end{align}
$Z$ depends on $\beta$  in the term $\exp(-\beta\phi(\pop))$ only, so differentiating with respect to $\beta$ we obtain
\begin{align}
\frac{\partial \log Z}{\partial \beta} & =  \frac{1}{Z} \sum_{\pop} -\phi(\pop) \Pu(\pop) \exp( -\beta \phi(\pop)) \nonumber \\
&= - \langle \phi(\pop) \rangle_{\pop\sim\Ps}\\
\intertext{Differentiating with respect to $\beta$ again,} 
\frac{\partial^2 \log Z}{\partial \beta^2} & = \frac{-1}{Z^2} \left( \sum_\pop -\phi(\pop) \Pu(\pop) \exp( -\beta \phi(\pop)) \right)^2 + \nonumber \\ 
& ~~~~~~~~~~~~~~~\frac{1}{Z} \sum_\pop \phi(\pop)^2 \Pu(\beta \pop) \exp( -\beta \pop) \\
&= -\langle \phi(\pop)^2 \rangle + \langle \phi(\pop) \rangle^2 \\
& = - \text{Var}_{\pop\sim\Ps}\{\phi( \pop) \} 
\end{align} 
Hence, for \emph{any} weight function $\phi$, for an evolutionary process with fixed population size, the following relation holds at stationarity: 
\begin{equation} 
\label{eq:fluctuationtheorem}
- \frac{\partial}{\partial \beta} \langle \phi(\pop) \rangle = \text{Var}\{\phi(\pop) \} 
\end{equation} 
Note that $\text{Var}\{\phi(\pop) \}$ is the variance of total fitness-weight of the population, not the variance of fitness within a population, nor the variance of fitness-weights of a genome drawn from the stationary distribution. This result is general, exact, and, we believe, novel.

We examine the consequences of this result in section \ref{EstimateEscape}. Here, we give an example of a bimodal stationary distribution, where $\frac{\partial}{\partial \beta} \langle \phi(\pop) \rangle$ diverges as $N\to\infty$, which corresponds to a critical value of $\beta$ at which there is a transition between two meta-stable regimes. 

\section{Stationary distribution in the infinite\\ population limit: the $\delta$-function method }\label{Secinfpop}

We now move on to exact calculation of stationary distributions in the infinite population limit, with an alternate presentation of the method developed and introduced in \cite{lember2020evolutionary}, section 6. We let the population size $N \to \infty$ while keeping the mutation rate $u = u_0 + u_1$ constant: we take the limit in this way because, from a biological point of view, we do not expect the mutation rate to vary with population size. Given the relationship $\alpha_0 = N \frac{u_0}{1-u}$, and the corresponding relationship for $\alpha_1$, it follows that to construct models with different population sizes but the same mutation rate $u$,  the concentration parameters $\alpha_0, \alpha_1$ must vary linearly with $N$. We will write:
\begin{align}
\begin{split}    \alpha_0^\prime &:= \frac{u_0}{1-u}\\
    \alpha_0(N)& := N\alpha_0^\prime
\end{split}
\begin{split}
    \alpha_1^\prime &:= \frac{u_1}{1-u}\\
    \alpha_1(N) &:= N \alpha_1^\prime
\end{split}  
\end{align}
In this section, $\alpha_0, \alpha_1$ depend on $N$, so that $u_0, u_1$ are constant, independent of $N$. Note that this variation of $\alpha$ with $N$ is in direct contrast with Bayesian modeling, in which the concentration parameters $\alpha_0, \alpha_1$ express a fixed prior belief that is independent of sample size. For notational simplicity, we will write $\vecalpha=(\alpha_0, \alpha_1)$, and $\vecu=(u_0, u_1)$.

We now examine the effect of letting $N \to \infty$ while keeping $\vecu$  constant in the case of a population with a single locus. As shown in \cite{lember2020evolutionary}, we may elegantly obtain the behavior of $\Ps$ as $N \rightarrow \infty$ by considering the posterior distribution of $\theta$.

Recall that, in the absence of selection, the distribution of the number of $1$'s at a locus is Beta-Binomial as given in equation \ref{eq:betabinomial}, and it is well known (see for example \cite{bernardo2009bayesian}) that this may be expressed as a continuous mixture of binomial distributions. Writing the binomial parameter as $\theta$, the prior distribution (i.e., the mixing distribution) over $\theta$ is the Beta distribution: 
\begin{equation}
    P(\theta \mid \alpha_0, \alpha_1) = \frac{1}{B(\alpha_0, \alpha_1)} (1-\theta)^{\alpha_0-1}\theta^{\alpha_1 -1}.\label{Ptheta}
\end{equation}
With $L$ loci, there are $L$ binomial parameters $\vectheta=(\theta_1, \ldots,\theta_L)$, and 
\begin{equation}
    P(\vectheta \mid \alpha_0, \alpha_1) = \left(\frac{1}{B(\alpha_0, \alpha_1)}\right)^L \prod_{j=1}^L (1-\theta_j)^{\alpha_0-1}\theta_j^{\alpha_1 -1} \label{Pvectheta}
\end{equation}
We will often omit explicit dependence on \(\vecalpha\) for clarity so that \(P(\vectheta)=P(\vectheta \mid \alpha_0, \alpha_1)\), \(P(x\mid\vectheta)=P(x\mid\vectheta, \alpha_0, \alpha_1)\), and so on.  

We then express the stationary distribution $\Ps$ in terms of $\vectheta$ as follows: 
\begin{align}
    1 &= \sum_{\pop\in\allpops} \Ps(\pop) \nonumber\\
    &= \frac{1}{Z(N)} \sum_{\pop\in\allpops}
    \Pu(\pop)f(\pop) \\
    &= \frac{1}{Z(N)} \sum_{\pop}
    \int_{\vectheta} f(\pop) P(\pop\mid\vectheta)
    P(\vectheta) d\vectheta \label{eq:thetalimit}\\
    &= \frac{1}{Z(N)} \int_{\vectheta}
    \left( \sum_{\pop} f(\pop) P(\pop\mid\vectheta)\right) P(\vectheta) d\vectheta \nonumber\\
    &= \frac{1}{Z(N)} \int_{\vectheta}
    E_{\pop\sim\vectheta}[f(\pop)] \, P(\vectheta) d\vectheta \\
\intertext{Note that the rows of $\pop$ are conditionally independent given $\vectheta$ so that the expectation of the product of row-fitnesses is equal to the product of expected row fitnesses so that we obtain:}
&= \frac{1}{Z(N)} \int_{\vectheta}
    E_{\bg\sim\vectheta}[f(\bg)]^N \, P(\vectheta) d\vectheta  \\
\intertext{expanding $P(\vectheta)$ using equation \ref{Pvectheta} and recalling \(\vecalpha=N\vecalpha^\prime\) gives}
&= \frac{1}{B(\alpha_0, \alpha_1)^L Z(N)} \int_{\vectheta}
    E_{\bg\sim\vectheta}[f(\bg)]^N \, 
    \prod_{j=1}^L (1-\theta_j)^{N\alpha_0^\prime -1} \theta_j^{N\alpha_1^\prime - 1} d\vectheta  \\
&= \frac{1}{B(\alpha_0, \alpha_1)^L Z(N)} \int_{\vectheta}
    \left( E_{\bg\sim\vectheta}[f(\bg)] \, 
    \prod_{j=1}^L (1-\theta_j)^{\alpha_0^\prime-\frac{1}{N}} \theta_j^{\alpha_1^\prime - \frac{1}{N}}\right)^N d\vectheta  
    \label{eq:Npowerdensity}
\end{align}
The posterior probability density of $\vectheta$ is, therefore:
\begin{equation}
\label{eq:posteriortheta} 
   P(\vectheta; f, N) = \frac{\left( E_{\bg\sim\vectheta}[f(\bg)] \, 
    \prod_{j=1}^L (1-\theta_j)^{\alpha_0^\prime-\frac{1}{N}} \theta_j^{\alpha_1^\prime - \frac{1}{N}}\right)^N}{B(\alpha_0, \alpha_1)^L Z(N)}
\end{equation}
where the denominator does not depend on $\vectheta$. 

To find the limiting posterior distribution of $\vectheta$ as $N\to\infty$, it is convenient to define a function $\maximiser$:
\begin{equation}
    \maximiser(\vectheta, f) := 
    E_{\bg\sim\vectheta}[f(\bg)] \, 
    \prod_{j=1}^L (1-\theta_j)^{\alpha_0^\prime} \theta_j^{\alpha_1^\prime}
\end{equation}
Considered as a function of $\vectheta$, $\maximiser$ is a continuous function defined on the compact region $[0,1]^L$. $\maximiser$ is zero on the boundary of $[0,1]^L$, and it is strictly positive in the interior. It follows that $\maximiser$ has a strictly positive supremum which is attained for some value $\vectheta=\vectheta^*$ in $(0,1)^L$. We may rewrite equation \ref{eq:posteriortheta} in terms of $\maximiser$
\begin{equation}
    P(\vectheta; f, N) =  \frac{ \maximiser(\vectheta, f)^N \prod_{j=1}^L \theta_j(1-\theta_j)}{B(\alpha_0, \alpha_1)^L Z(N)}
\end{equation}
For simplicity of exposition, let us only consider the case where $\maximiser$ has a \emph{unique} global maximum value so that $\vectheta^*$ is unique. 
First, we show that 
\begin{equation}
    \lim_{N\to\infty} \argmax_{\vectheta} P(\vectheta \mid f, N) = \argmax_{\vectheta} \maximiser(\vectheta,f)
\end{equation}
To show this, note that 
\begin{align}
    \argmax_{\vectheta} P(\vectheta \mid f, N) &= \argmax_{\vectheta} P(\vectheta \mid f, N)^\frac{1}{N}\\
    &= \argmax_{\vectheta} \frac{ \maximiser(\vectheta,f)^N \prod_{j=1}^L (1-\theta_j)^{\frac{1}{N}}\theta_j^{\frac{1}{N}} }{B(\alpha_0, \alpha_1)^L Z(N)}\\
    \intertext{and since $\lim_{N \to. \infty}(1-\theta)^{\frac{1}{N}}\theta^{\frac{1}{N}}=1$ for $0<\theta<1$, we have}
    &= \argmax_{\vectheta} \maximiser(\vectheta,f)
\end{align}
as required. 

From equation \ref{eq:Npowerdensity}, it is evident that the limiting form of the posterior of $\vectheta$ is a delta function at $\vectheta^*$; this is because the integrand is non-negative, it integrates to 1, and we have assumed a unique global maximum.

Using the fact that the limiting posterior of $\vectheta$ is a delta function at $\vectheta^*$, we can see from equation \ref{eq:thetalimit} that the marginal distribution of genomes in the population in the infinite $N$ limit is :
\begin{equation}
    P(\bg) = \frac{1}{Z_\infty} P(\bg \mid \vectheta^*) f(\bg)
\end{equation}
where 
\begin{equation}
    Z_\infty(f,\vecalpha) = \sum_{\bg\in\{0,1\}^L} P(\bg \mid \vectheta^*) f(\bg)
\end{equation}
In conclusion, we can calculate the distribution of genomes in the infinite population using the following computational method. 
First, given $f$,  we implement a function 
\begin{align}
    F(\vectheta, \vecalpha) & := E_{\bg\sim\vectheta}[f(\bg)]\\
   &  = \sum_{\bg\in\{0,1\}^L} f(\bg)\prod_{j=1}^L \left( [\bg(j)=0](1-\theta^*_j) + [\bg(j)=1] \theta^*_j \right)
\end{align}
where we use the notation $[ expression]$ to equal 1 if $expression$ is true, and $0$ if $expression$ is false. 
This is a canonical but usually inefficient method of computing $ F(\vectheta, \vecalpha)$ by summing over all $2^L$ possible genomes. In specific cases outlined in section \ref{FF}, there are more efficient ways of computing $F$. $F$ is then used in computing $\maximiser$.

Next, we use numerical optimization to find the $\vectheta^*$ that maximizes $\maximiser$. Noting that
\begin{equation}
    Z_\infty(f,\vecalpha) = F(\vectheta^*, \vecalpha)
\end{equation}
we obtain the infinite-population stationary genome probabilities as 
\begin{align} 
\Ps(\bg) & = \frac{1}{Z_\infty}P(\bg\mid\vectheta^*) f( \bg )\\
&= \frac{1}{F(\vectheta^*, \vecalpha)} f(\bg) \prod_{j=1}^L \left( [\bg(j)=0](1-\theta^*_j) + [\bg(j)=1] \theta^*_j \right)
\label{eq:infpopfitnessreweighting}
\end{align}
As far as we know, this approach to computing the infinite population stationary distribution does not appear in the literature. Next, we demonstrate this method on four different fitness functions. 

\section{Four fitness functions and their stationary \\ 
distributions}\label{FF}
\label{sec:fourfitnessfunctions}

We consider four different fitness functions for multilocus genomes. For each of them, we demonstrate how to calculate the stationary distribution for both finite populations and the infinite population limit. 





\subsection{Perfect fitness} 
`Perfect fitness' is defined as:
\begin{equation}
   \fperfect(\bg)=
\begin{cases}
1, \: {\textsf{  if\,}} \bg(1)=\cdots=\bg(L)=1\\
e^{-\beta}, \: {\textsf{  otherwise}}.\\
\end{cases}
\end{equation}
This fitness is referred to in the literature as the ``needle-in-a-haystack'' problem \cite{goldberg1991genetic} or the ``sharp peak'' landscape \cite{alves1996population}.

\subsubsection{Perfect fitness: infinite population}
To implement the $\delta$-function method of section \ref{Secinfpop}, we need to implement a function that computes 
\begin{equation*}
    F^{\perfect}(\vectheta) = E_{\bg\sim\vectheta}[ \fperfect(\bg)]\\
\end{equation*}
The simplest and canonical way to do this is to sum over all $2^L$ genomes:
\begin{equation}
    F^{\perfect}(\vectheta) = \sum_{\bg\in\{0,1\}^L} \fperfect(\bg)\prod_{i=1}^{L} \theta_i^{\bg(i)}(1-\theta_i)^{1-\bg(i)}
\end{equation}
However, since there is only one exceptional fit genome, a more efficient computation for perfect fitness is: 
\begin{align}
    F^{\perfect}(\vectheta) &= \prod_{i=1}^L\theta_i + e^{-\beta}(1 - \prod_{i=1}^L\theta_i)\\
    &= e^{-\beta} + (1-e^{-\beta})\prod_{i=1}^L\theta_i
\end{align}
To find $\vectheta^*$, we find
\begin{equation}
    \vectheta^* = \argmax_{\vectheta} ~ F^{\perfect}(\vectheta) \prod_{j=1}^L (1-\theta_j)^{\alpha_0^\prime} \theta_j^{\alpha_1^\prime}
\end{equation}
where $\maximiser(\vectheta,f^{\textsf{  perfect}})=F^{\perfect}(\vectheta) \prod_{j=1}^L (1-\theta_j)^{\alpha_0^\prime} \theta_j^{\alpha_1^\prime}$. For this fitness function,  $\vectheta^*=(\theta^*_1,\ldots,\theta^*_L)$  which maximises $\maximiser(\vectheta, f^{\textsf{  perfect}})$ has all elements equal: that is $\theta^*_1=\cdots=\theta^*_L$.

To prove this, let us suppose the contrary, that there is some $\vectheta$ for which $\maximiser(\vectheta,f^{\textsf{  perfect}})$ is a global maximum, and such that some elements of $\vectheta$ are unequal: specifically that there are  $i$, $j$ such that $\theta_i \neq \theta_j $. We now establish a contradiction, by exhibiting $\vectheta^\prime$ such that $\maximiser(\vectheta^\prime, f^{\textsf{  perfect}}) > \maximiser(\vectheta, f^{\textsf{  perfect}})$. Let $\theta^\prime_k = \theta_k$ for all $k\neq i,j$ (note we do not assume all $\theta_k$ are the same, although they can be), and let $\theta^\prime_i=\theta^\prime_j=\sqrt{\theta_i\theta_j}$.

Then, since $F^{\textsf{  perfect}}(\vectheta)$ depends only on $\theta_1\times\cdots\times\theta_L$, it is clear that $F^{\textsf{  perfect}}(\vectheta^\prime)=F^{\textsf{  perfect}}(\vectheta)$, so that 
\begin{align}
    \frac{\maximiser(\vectheta^\prime,f^{\textsf{  perfect}},\vecalpha)}{\maximiser(\vectheta,f^{\textsf{  perfect}},\vecalpha)} & = 
    \frac{(\sqrt{\theta_i\theta_j})^{2\alpha_1} (1- \sqrt{\theta_i\theta_j})^{2\alpha_0}}{
    (\theta_i)^{\alpha_1}(1-\theta_i)^{\alpha_0} (\theta_j)^{\alpha_1}(1-\theta_j)^{\alpha_0}}\\
    & =  \frac{(1- \sqrt{\theta_i\theta_j})^{2\alpha_0}}{
    (1-\theta_i)^{\alpha_0} (1-\theta_j)^{\alpha_0}}\\
    & > 1 ~~~ \textrm{iff $\frac{(1- \sqrt{\theta_i\theta_j})^{2}}{
    (1-\theta_i) (1-\theta_j)}$}>1
\end{align}
The final inequality follows from Jensen's inequality in the following way. Note first that $\log(1-\theta)$ is concave in $\theta$ for $0 < \theta < 1$. Let $x=\log(\theta)$; then $\log(1-\theta)= \log(1-e^x)$ is also concave in $x$, since $x$ is a monotonic function of $\theta$. We now apply Jensen's inequality in $\log(1-e^x)$ vs. $x$. Putting $x_i=\log(\theta^*_i)$ and $x_j = \log(\theta^*_j)$, by Jensen's inequality, 
\begin{align}
    \log(1-e^{\frac{x_1+x_2}{2}}) & > \frac{\log(1-e^{x_1}) + \log(1-e^{x_2})}{2}\\
\intertext{Exponentiating and squaring gives}
    (1-e^{\frac{x_1+x_2}{2}})^2 & > (1-e^{x_1})(1-e^{x_2})\\
\intertext{and changing variables from $x_1, x_2$ to $\theta_i, \theta_j$ gives}
    (1-\sqrt{\theta_i\theta_j})^2 & > (1-\theta_i)(1-\theta_j).
\end{align}
We, therefore, establish that all elements of a globally optimal $\vectheta^*$ must be equal. 
It follows that we can find $\vectheta^*= (\theta^*, \ldots, \theta^*)$ by a maximisation in one variable only:
\begin{equation}
\theta^*  = \argmax_\theta \left(\theta^{\alpha^\prime_1}(1-\theta)^{\alpha^\prime_0}\right)^L \, (e^{-\beta} + (1 - e^{-\beta})\theta^L)\\
\end{equation}

\subsubsection{Perfect fitness: finite population}
To calculate the stationary distribution of the number of perfect genomes in a finite population, we first calculate the stationary distribution assuming no selection. Then we will re-weight each population's probability according to the number of perfect genomes it contains. 

It will be convenient to define the function $r_k$, which counts the number of genomes in a population $\pop$ that have a prefix of $k$ or more 1s: 
\begin{equation}
r_k(\pop) := \#\{ i : \prefix(\bg_i) \ge k \} 
\end{equation} 
for $0 \le k \le L$, where $\pop \in \{0,1\}^{N\times L}$. Note that $N = r_0(X) \ge r_1(X) \ge \cdots \ge r_L(X) \ge 0$.  $r_L(\pop)$ is the number of fully perfect genomes in $\pop$ with all elements equal to 1.


Let us define the following function, for which we will give a recursive definition. For a population $\pop$ with $N$ genomes and $L$ loci, without selection, and for given $\alpha$, let the probability that $k$  genomes out of $N$ are perfect be denoted: 
\begin{equation} 
\Pperfect(N, L, k) := \sum_{\pop \in \setzeroone^{N\times L}} \Pu(\pop) [ r_L(\pop) = k ] 
\end{equation} 
where we use the notation $[ expression]$ to equal 1 if $expression$ is true, and $0$ if $expression$ is false. 

First, note that for a population with only one locus, for given $\vecalpha$, the distribution of the number of 1s is Beta-Binomial: 
\begin{align} 
\label{eq:perfectbasecase} 
\Pperfect(N, 1, k) &= \bebi( n_1; N, \alpha_0, \alpha_1)\\
&= {N \choose n_{1}}\frac{(\alpha_0)_{N-n_1}(\alpha_1)_{n_{1}}}{(\alpha_{0} + \alpha_1)_{N}}
\end{align} 
The following recursion holds for $N, L_1, L_2 >  0, N \ge m \ge 0$, and in conjunction with the previous equation\ \ref{eq:perfectbasecase} it can be used to compute $\Pperfect$ for all argument values: 
\begin{equation} 
\label{eq:perfectrecursion} 
\Pperfect(N,L_1 + L_2, m) = \sum_{k=m}^N \Pperfect(N, L_1, k) \Pperfect(k, L_2, m) 
\end{equation} 
The recursion follows by the following argument. Consider the first $L_1$ columns of $\pop$; the probability that $k$ of the genomes are perfect for these $L_1$ columns --- that is, the probability that exactly $k$ out of these $N$ genomes have prefixes consisting of $L_1$ or more 1s --- is $\Pperfect(N,L-1,k)$. Now consider the first $L_1 + L_2$ columns of $\pop$. Any genome that is perfect for the first $L_1 + L_2$ columns must necessarily be perfect for the first $L_1$ columns; therefore, we need only consider the continuations of the $k$  genomes that are perfect for the first $L_1$ columns. By exchangeability of the rows of $\pop$, we may sample these $k$ rows of columns $L_1+1$ to $L_1 + L_2$ first, and the probability that exactly $m$ of these $k$ rows with $L_2$ columns are perfect is $\Pperfect(k,L_2,m)$.
Hence to compute $\Pperfect(N,L_1+L_2,m)$, we must sum over the distinct cases where $m, m+1, \ldots, N$ rows of 
the first $L_1$ columns are perfect, so the recursion follows. 

The probability $\Psperfect(N,L,k)$ that there are $k$ out of $N$ perfect genomes in a selected population is then easily given in terms of $\Pperfect$. First, define the un-normalized probability times fitness for $k$ perfect out of $N$:  
\begin{equation}
\Zperfect(N,L,k) := \exp(-\beta (N-k)) \Pperfect( N, L, k)
\end{equation}
and the normalization factor for the whole fitness-weighted distribution is
\begin{equation}
\Zperfect(N,L) := \sum_{k=0}^N \exp(-\beta (N-k)) \Pperfect(N,L,k) = \sum_{k=0}^N \Zperfect(N,L,k) 
\end{equation}
so that the normalized stationary probability is
\begin{equation}
\Psperfect( N, L, k ) = \frac{\Zperfect(N,L,k)}{\Zperfect( N, L ) }
\end{equation} 

\noindent Evidently, we can readily compute $\Psperfect$ for any $\beta$ if we can compute $\Pperfect(N,L,k)$.



\subsection{Prefix fitness}

Given a binary sequence $\bg$, define $\prefix(\bg)$ to be the length of the longest prefix of $\bg$ that consists entirely of ones. For example, $\prefix(1110101)=3$, and $\prefix(01111) = 0$. When the prefix length is used to determine fitness, this is called the \emph{leading ones} problem \cite[Section 3.3.6]{Doerr:2020:Complexity}.

The `prefix fitness' of a genome $\bg$ of length $L$ is defined as
\begin{equation} 
\label{eq:fprefixdef}
\fprefix(\bg) = \exp( \beta ~ \prefix(\bg) )
\end{equation} 
 If $\prefix(\bg)=0$, $\fprefix(\bg)=1$; if $\prefix(\bg)=L$, then $\fprefix(\bg) = e^{\beta L }$. 
 
\subsubsection{Prefix fitness: infinite population}
Note that for $0 \le k < L$,
\begin{equation}
    P(\prefix(\bg)=k \mid \vectheta ) = (1-\theta_{k+1}) \, \prod_{j=1}^k \theta_j
\end{equation}
and 
\begin{equation}
    P( \prefix(\bg)=L \mid \vectheta \,)=\prod_{j=1}^L \theta_j
\end{equation}
The expected fitness function needed is
\begin{align} 
F^{\prefix}(\vectheta) = \sum_{k=0}^L P(\prefix(\bg)=k \mid \vectheta)\exp(\beta k)
\end{align} 
Recall that $\maximiser(\vectheta,f^{\textsf{  prefix}})=F^{\prefix}(\vectheta) \prod_{j=1}^L (1-\theta_j)^{\alpha_0^\prime} \theta_j^{\alpha_1^\prime}$.
To find $\vectheta^*$, we find
\begin{equation}
    \vectheta^* = \argmax_{\vectheta} \maximiser( \vectheta, \fprefix, \vecalpha).
\end{equation}
The stationary distribution is then obtained by fitness-reweighting the product of binomial distributions with parameters $\vectheta$, according to equation \ref{eq:infpopfitnessreweighting}.

\subsubsection{Prefix fitness: finite population}
To compute the stationary distribution with leading-ones (prefix) fitness for populations of size $N\times L$, we may use the following recursion on fitness-weighted probability. It is convenient to define the following function: 
\begin{equation}
    \Zprefix(n, l, k) := \sum_{\pop\in\{0,1\}^{n\times l}} \Pu(\pop) \fprefix(\pop)[r_l(\pop)=k]
\end{equation}
where $r_l(\pop)$ is the number of elements of the $n\times l$ population $\pop$ that have prefixes of length $l$ or more, and $[r_l(\pop)=k]$ equals 1 if the statement is true and zero if it is false. In other words,  $\Zprefix(n,l,k)$ is the unnormalized fitness-weighted breeding probability for populations with exactly $k$ prefixes of length $l$ or more. The normalizing factor for the stationary distribution  for a population of $N$ genomes of length $L$ is then: 
\begin{equation} 
\Zprefix(N,L) = \sum_{\pop\in\{0,1\}^{N\times L}} \Pu(\pop) \fprefix(\pop) = \sum_{k=0}^N \Zprefix(N,L,k) 
\end{equation}
We give a recursion for computing $\Zprefix(N,L,k)$; the recursion is over all of $N$, $L$, and $k$.
\begin{equation} 
\Zprefix(N,1,k) = \bebi(k; N,\alpha_0, \alpha_1) \exp(\beta k)
\end{equation} 
and for $l_1, l_2 \ge 1$, $l_1 + l_2 \le L$, 
\begin{equation} 
\Zprefix(N, l_1 + l_2, m) = \sum_{k=m}^N \Zprefix(N,l_1, k) \Zprefix(k,l_2,m) 
\end{equation} 
and the normalizing factor is 
\begin{equation} 
\Zprefix(N,L) = \sum_{k=0}^N \Zprefix(N,L,k)
\end{equation} 
so that
\begin{equation}
\Psprefix(N,L,k) = \frac{ \Zprefix(N,L,k) }{ \Zprefix(N,L) } 
\end{equation} 

The recursive (or dynamic programming) computation of $\Zprefix(N,L)$ yields a table of values of $\Zprefix(n,l,k)$ for $1\le n\le N$, $1\le l\le L$, and $0\le k \le n$. Many summary statistics of the stationary distribution for prefix fitness can be calculated using this table. In particular, the marginal distribution of the number of valid prefixes (not necessarily completed) at the $l$'th locus is given by: 
\begin{equation} 
\Psprefix( \text{$k$ valid prefixes at locus $l$} ) = \frac{ \Zprefix(N,l,k) \sum_{m=0}^k \Zprefix(k,L-l,m)} {\Zprefix(N,L)} 
\end{equation} 
We define the number of valid prefixes at locus zero to be $N$, and the number of valid prefixes at locus 1 equals the number of 1s at locus 1. 

\subsection{Sum fitness} 

In Sum fitness (close to a multiplicative fitness model \cite{lewontin1964interaction} or the ONEMAX landscape \cite{hesser1990towards}), the key to the analysis is that the stationary distribution at each locus is independent of the distributions at other loci, and we establish this first. Sum fitness of a single genome of length $L$ is defined as 
\begin{equation}
    \fsum(\bg) := \exp\left( \beta \sum_{j=1}^L \bg(j) \right),
\end{equation}
and for a population $\pop\in\{0,1\}^{N\times L}$
\begin{align}
    \fsum(\pop) & = \prod_{i=1}^N\fsum(\bg_i)\\
    & = \exp\left( \beta \sum_{1\le i\le N, 1\le j\le L} \pop_{ij}\right)\\
    \intertext{ and expressing it as a product of functions of columns (loci) of $\pop$ we get}
    & = \prod_{1\le j \le L } \exp{\left( \beta 
    \sum_{1\le i \le N}\locus_j(i) \right)}
\end{align}
Note: \(\fsum(\pop)\) is not to be taken as the mean fitness of the population, which is \(\Zsum(N,1)\) to follow.

It follows that the stationary distribution is the product of independent column-distributions: 
\begin{align}
\Pssum(\pop) &= \frac{1}{\Zsum(N,L)} \Pu(\pop) \fsum(\pop)\\
& =  \prod_{j=1}^L \frac{1}{\Zsum(N,1)} \Pu(\locus_j)\exp\left(\beta \sum_i\locus_j(i)\right)
\end{align}
where as usual $\Zsum(N,L)=\Zsum(N,1)^L$ --- the population mean fitness --- is the normalising factor for a population of size $N$ with $L$ loci, and specifically
\begin{equation}
    \Zsum(N,1) = \sum_{k=1}^N {N \choose k} \frac{(\alpha_0)_{N-k} (\alpha_1)_k}{(\alpha_0+\alpha_1)_N}e^{-\beta (N-k)}
\end{equation}
This column-independence holds for finite populations and, therefore, in the infinite population limit. Note that although
columns (loci) are independent, rows (genomes) within a population are not independent. 

\subsubsection{Sum fitness: infinite population} 
Since loci are independent, we need consider only the case $L=1$, and we need to optimize for only a single parameter $\theta$. The expected fitness function is: 
\begin{equation}
    F^{\sfsum}(\theta) = \sum_{k=0}^L {L \choose k} (1-\theta)^{k}\theta^{L-k} e^{-\beta k}
\end{equation}
Recall that $\maximiser(\vectheta,f^{\textsf{  sum}})=F^{\sfsum}(\vectheta) \prod_{j=1}^L (1-\theta_j)^{\alpha_0^\prime} \theta_j^{\alpha_1^\prime}$.
To find $\vectheta^*$, we find
\begin{equation}
    \vectheta^* = \argmax_{\vectheta} \maximiser( \vectheta, \fsum, \vecalpha).
\end{equation}
An explicit expression for $\theta^*$ is given in \cite{lember2020evolutionary}.

\subsubsection{Sum fitness: finite population}
For a finite population with $L=1$ --- that is, only one locus --- the stationary distribution of the number of 1s follows from 
\begin{equation} 
\Pssum( \text{$k$ 1s at locus $l$} ) = \frac{ \bebi(k;N,\alpha_0, \alpha_1) e^{\beta k}} {\Zsum(N,1)} 
\end{equation} 
For populations with $L>1$, the stationary distribution is the $L$-fold direct product of the stationary distribution given above for a single locus.

\subsection{One Error fitness}

One Error fitness is defined as 
\begin{equation}
    f^{\textsf{  OneError}}(\bg) = \begin{cases}
    1, \: \text{if $\sum_{j=1}^L \bg(j) \ge L-1$}\\
    e^{-\beta} \: \textsf{otherwise.} 
    \end{cases}
\end{equation}
That is, a genome is fit if it contains at most one zero, and it is unfit if it contains more than one zero. 
\subsubsection{OneError fitness: infinite population} 

We once again use the method outlined in \ref{Secinfpop}.
\begin{equation}
    F^{\textsf{  OneError}}(\vectheta) := e^{-\beta} + (1-e^{-\beta})\left(\prod_{j=1}^L\theta_j\right)\left(1 + \sum_{i=1}^L\frac{1-\theta_i}{\theta_i}\right)
\end{equation}
Recall that $\maximiser(\vectheta,f^{\textsf{  oneerror}})=F^{\oneerror}(\vectheta) \prod_{j=1}^L (1-\theta_j)^{\alpha_0^\prime} \theta_j^{\alpha_1^\prime}$.
To find $\vectheta^*$, we find
\begin{equation}
    \vectheta^* = \argmax_{\vectheta} \maximiser( \vectheta, \foneerror, \vecalpha).
\end{equation}

\subsubsection{OneError fitness: finite population}

\begin{figure}
    \centering
    \includegraphics[scale=0.3]{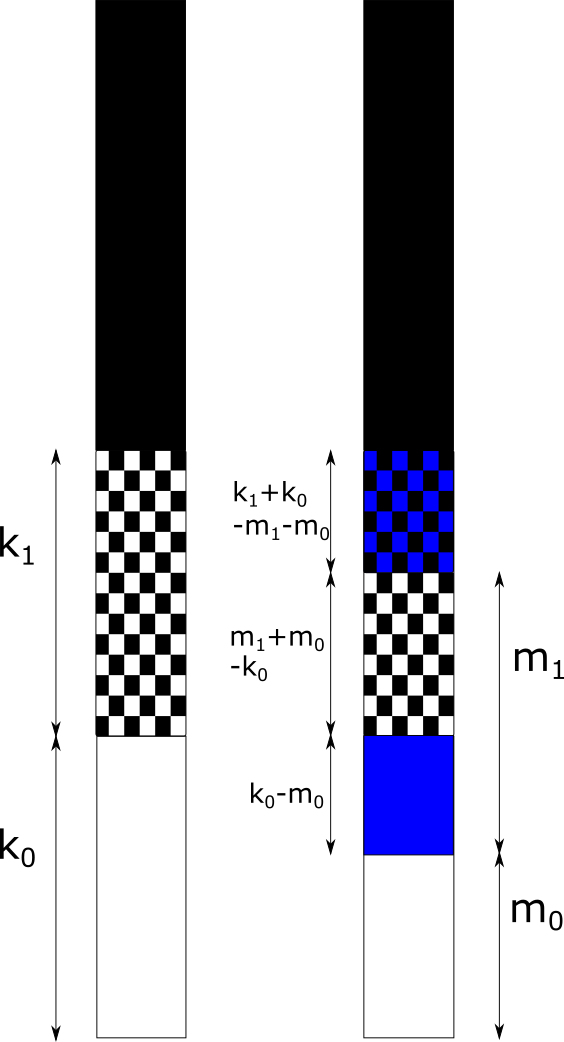}
    \caption{Here, we visualize the recursion required to calculate the number of genomes with a single error based on the urn probabilities. The probability of an urn of size \(n\) containing \(n_{0}\) white balls and \(n_{1}\) black balls is \(\bebi(n_{1}; n_{0}+n_{1})\). We show here an urn representing each genome's \(j\)th locus. There are \(k_{0}\) genomes that contain only white loci up until (and including) locus \(j\), and \(k_{1}\) which have only accumulated a single error at any point on the genome up until locus \(j\). We want to calculate how many genomes will be perfect or have a single error up until locus \(j+1\). Of the \(k_{0}\) perfect genomes, \(m_{0}\) will remain perfect up until the next site, the \(k_{0}-m_{0}\) remaining will become one error genomes. If we assume that the total number of genomes with a single error at locus \(j+1\) is \(m_{1}\), then of the \(k_{1}\) genomes with a single error, \(m_{1}-(k_{0}-m_{0})\) must fail to gain a further error, leaving a set of \(k_{1}+k_{0}-m_{1}-m_{0}\) genomes which have gained a second error. Thus out of the total of \(k_{0}+k_{1}\) genomes of interest, \(2m_{0}+m_{1}-k_{0}\) must gain a white ball at urn \(j+1\) and \(k_{1}+2k_{0}-m_{1}-2m_{0}\) must gain a black ball.This occurs with probability \(\bebi(2m_0 + m_1 - k_0;k_0+k_1)\). There are then additional combinatorial factors for distributing the genomes into the groups above.}
    \label{iterator}
\end{figure}

To calculate the stationary distribution of the number of perfect or single error genomes in a finite population, we first calculate the stationary distribution assuming no selection. Then we will re-weight the probability of each population according to the number of perfect or single error genomes it contains. 

Recall the function $r_k$ which counts the number of genomes in a population $\pop$ that have a prefix of $k$ or more 1s: 
\begin{equation}
r_k(\pop) := \#\{ i : \prefix(\bg_i) \ge k \}. 
\end{equation} 
Here, we are interested in the number of genomes with a prefix of length \(k\) which contain one or fewer \(0\)'s:
\begin{equation}
s_k(\pop) := \#\{ i : \sum_{i=1}^{k} \bg_i \geq k-1 \}.
\end{equation} 

Let us define the following function, for which we will give a recursive definition. For a population $\pop$ with $N$ genomes and $L$ loci, without selection, and for given $\alpha$, let the probability that $m_0$  genomes out of $N$ are perfect and $m_1$ genomes out of $N$ have a single error be denoted: 
\begin{equation} 
\Poneerror(N, L, m_0,m_1) := \sum_{\pop \in \setzeroone^{N\times L}} \Pu(\pop) [ s_L(\pop) = m_0+m_1 \land r_L(\pop) = m_0] 
\end{equation} 
where we use the notation $[ expression]$ to equal 1 if $expression$ is true, and $0$ if $expression$ is false.

We now derive a recursive expression for $\Poneerror(N,L,m_0,m_1)$. In the base case, $L=1$; each genome is either 0 or 1, so 
\begin{equation}
    \Poneerror(N,1,m_0,m_1) = \bebi(m_0;N)
\end{equation}
Next, suppose that we have computed the values $\Poneerror(N, L, k_0, k_1)$ for all $k_0+k_1\le N$. We wish to derive expressions for $\Poneerror(N, l+1, m_0, m_1)$ in terms of values of $\Poneerror(N, l, k_0, k_1)$.

Consider the situation depicted in figure \ref{iterator}, where $r_l(\pop)=k_0$ and $s_l(\pop)=k_1+k_0$. Note that any perfect $l+1$-prefix must have a perfect $l$-prefix, followed by a 1 at locus $l+1$. Any one-error $l+1$-prefix must consist of either a perfect $l$-prefix followed by a zero at $l+1$ or a one-error $l$-prefix, followed by a 1 at $l+1$. Let the number of perfect $l+1$-prefixes be $m_0$, and the number of one-error $l+1$-prefixes be $m_1$. There must be exactly $m_0 \le k_0$ ones matching the $k_0$ perfect $l$-prefixes. The number of one-error $l+1$-prefixes is $k_0-m_0$ cases where a 0 at $l+1$ matches a perfect $l$-prefix, and $m_0+m_1-k_0$ cases where a 1 at locus $l+1$ matches a one-error $l$-prefix. Referring to figure \ref{iterator}, it follows that there must be $2m_0 + m_1 - k_0$ ones in the $k_0 + k_1$ relevant elements of locus $l+1$ The total number of configurations of $2m_0 + m_1 - k_0$ ones distributed among $k_0 + k_1$ rows is $${ {k_0+k_1} \choose {2m_0 + m_1 - k_0}},$$ but only
$${ k_0 \choose  m_0}{ k_1 \choose {m_0+m_1-k_0}}$$ of these configurations will yield $m_0$ perfect $l+1$-prefixes and $m_1$ one-error $l+1$-prefixes. Hence the probability of $m_0$ perfect prefixes and $m_1$ one-error prefixes at $l+1$, given $k_0$ perfect and $k_1$ one-error prefixes at $l$, may be written: 

\begin{multline}
   \Poneerror(m_{0},m_{1}|k_{0},k_{1}) :=\\
   \bebi(2m_0 + m_1 - k_0;k_0+k_1) 
   \frac{\binom{k_{0}}{m_{0}}\binom{k_{1}}{m_{1}+m_{0}-k_{0}}}{\binom{k_{1}+k_{0}}{2m_0 + m_1 - k_0}},
\end{multline}



The complete recursion is obtained by summing over all possible values of \(k_{0}\) and \(k_{1}\), and multiplying by the factors $\Poneerror(N,l,k_0,k_1)$:
\begin{multline}
    \Poneerror(N,l+1,m_{0},m_{1})= \\ 
    \sum_{k_{0}=m_0}^{m_{0}+m_1}\sum_{k_{1}=m_{0}+m_1 - k_0}^{n-k_0}
    \Poneerror(N,l,k_{0},k_{1})\Poneerror(m_0,m_1\mid k_0, k_1).
\end{multline}

The probability $\Psoneerror(N,L,k_0,k_1)$ that there are $k_0$ out of $N$ perfect genomes and \(k_1\) out of \(N\) genomes with a single \(0\) in a selected population is then easily given in terms of $\Poneerror$. First, define the un-normalised probability times fitness for $k_0$ perfect and $k_1$ one error out of $N$:  
\begin{equation}
\Zoneerror(N,L,k_0,k_1) := \exp(-\beta (N-k_0-k_1)) \Poneerror( N, L, k_0,k_1)
\end{equation}
and the normalization factor for the whole fitness-weighted distribution is
\begin{equation}
\Zoneerror(N,L) :=\sum_{k=0}^N \sum_{k_1=0}^{N-k_0} \Zoneerror(N,L,k_0,k_1) 
\end{equation}
so that the normalized stationary probability is
\begin{equation}
\Psoneerror( N, L, k_0, k_1 ) = \frac{\Zoneerror(N,L,k_0,k_1)}{\Zoneerror( N, L ) }
\end{equation} 

\section{Calculations}

We will now demonstrate exact calculations of the stationary distribution for finite and infinite populations and show surprising differences in the responses to selection between fitness functions. We then consider a deeper question: could error-correcting codes provide a better evolutionary response than naive encodings? To answer this question, we compare the theoretical performance of naive encodings with Hamming codes, demonstrating that Hamming codes yield substantially higher mean fitness under certain conditions. As far as we know, this is the first demonstration that error-correcting codes could improve the precision of adaptation. Although our example is artificial, the fact that error correction can, in principle, improve the precision of adaptation raises many further questions about genetic architecture, which are beyond the scope of this paper. 

\subsection{Convergence of finite population distribution to infinite population limit}

\begin{figure}[t]
    \centering
    \includegraphics[scale=0.5]{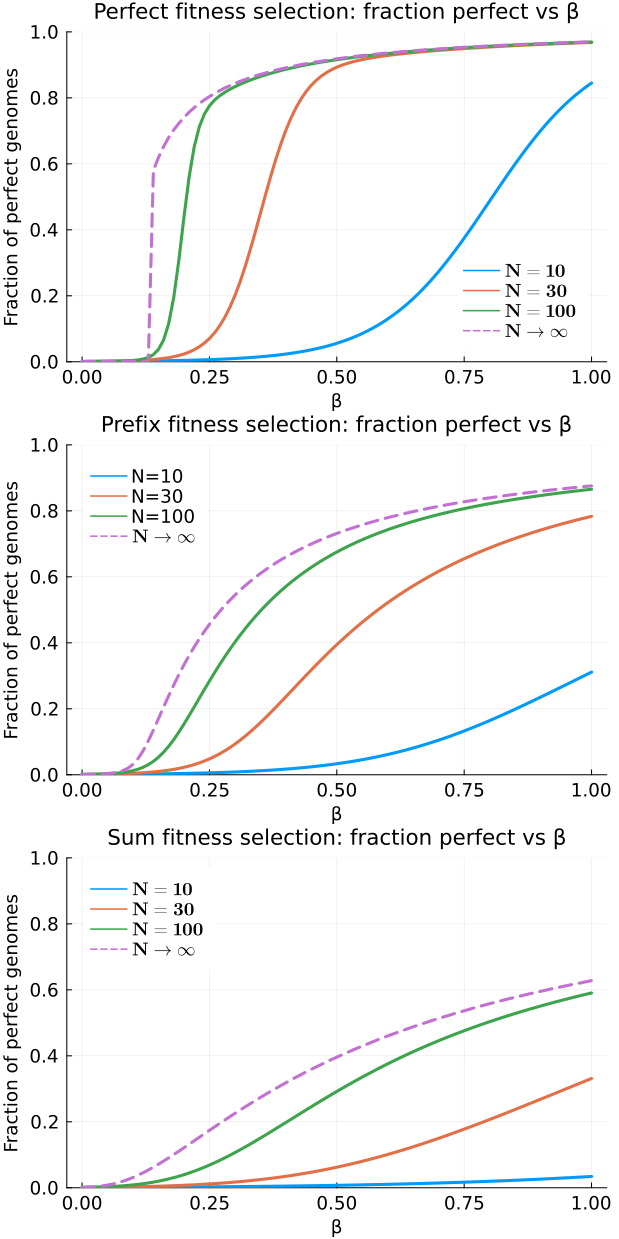}
    \caption{These graphs show convergence of finite-population statistics (calculated recursively) to the infinite population limit (calculated using the $\delta$-function method). The expected fraction of perfect (all-1s) genomes is plotted against selection intensity $\beta$, under selection using perfect, prefix, and sum fitness. Mutation rate $u=0.01$, and genome length $L=10$ for all curves. Population sizes are 10, 30, 100, and the infinite limit. For each fitness function, the maximum possible fitness is 1, and the minimum possible is $e^{-\beta}$. $\beta$ is shown on the x-axis in each graph. The top graph for perfect fitness shows a discontinuous jump in the fraction of perfect genomes as $\beta$ increases in the infinite population limit. The lower graphs of selection with prefix and sum fitness show substantially smaller fractions of perfect genomes.}
    \label{FiniteInfinite}
\end{figure}

We have shown two methods to calculate the stationary distributions for finite populations and the infinite population limit. Figure \ref{FiniteInfinite} shows plots of the theoretical fraction of perfect genomes versus $\beta$ for the four fitness functions, with separate curves for different population sizes, including the infinite population limit. All four graphs show the finite population statistics converging rapidly to the infinite population statistics: the curves for $N=100$ are already close to the curve for $N\to\infty$. 

We vary the values of $\alpha_0, \alpha_1$ for each population size $N$ so that the mutation rates are the same for all population sizes, at $u_0=u_1=0.05$. In addition, for each fitness function, the selection intensity parameter $\beta$ in the formulae in section \ref{FF} has been normalized to make all four fitness functions comparable. For all four fitness functions, the maximal fitness is 1, and the minimal fitness is $e^{-\beta}$. This rescaling makes the four fitness functions directly comparable.

\subsection{Perfect fitness: metastability and a bound on escape time}
\label{EstimateEscape}

An advantage of exact calculations of stationary distributions is to gain insight into multi-modal distributions, which are hard to characterize by simulations. This difficulty is because different modes of the stationary distribution correspond to metastable regimes of the simulation. Transitions between the metastable modes may be rare, so infeasibly long simulations are needed to observe these transitions.

\begin{figure} 
   \centering
   \includegraphics[scale=0.3]{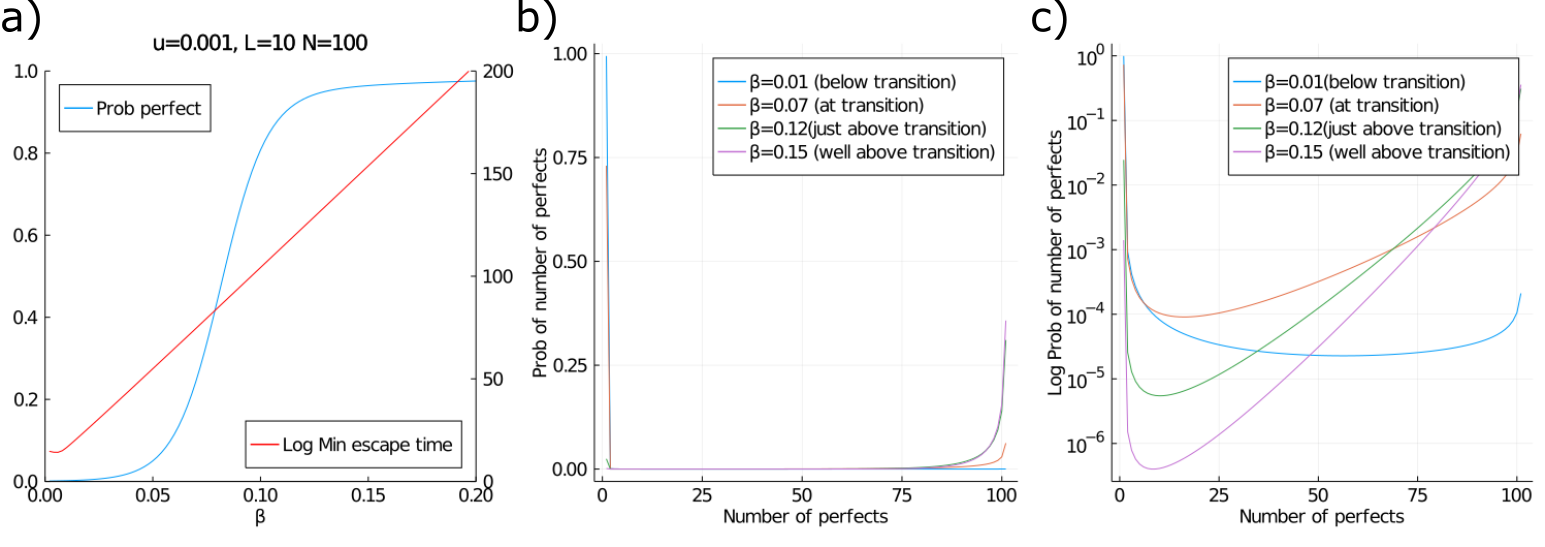}
 \caption{Graph a) shows the fraction of perfect genomes in the stationary distribution versus $\beta$, for $N=100, L=10$, and $u=0.001$. For sufficiently weak selection (low $\beta$), the fraction of perfect genomes is close to zero, rising to 1 for high $\beta$. Graph b) shows histograms, and c) log-histograms, of the fraction of perfect genomes in populations sampled from the stationary distribution for 4 selected values of $\beta$. For all values of $\beta$, the distribution is bimodal, with maxima at 0 and 1. With perfect fitness, the population is either nearly perfect or else almost wholly `imperfect', and both regimes are metastable. The rate of transition between modes is bounded below by the reciprocal of the smallest probability in the histogram: this lower bound on the log escape time is shown by the red line in a).}
    \label{PerfectFitnessHistograms}
\end{figure}

Using our recursions, it is straightforward to calculate the expected histogram of the number of perfect genomes (with no `bad' alleles) in the population at stationarity. Figure \ref{PerfectFitnessHistograms}c) shows this histogram for four values of $\beta$. Note that the relative heights of the modes depend strongly on $\beta$. Figure \ref{PerfectFitnessHistograms}a) shows that for sufficiently weak selection (low $\beta$), the fraction of perfect genomes is close to zero, but it rises to 1 for high $\beta$. However, we see in figures \ref{PerfectFitnessHistograms}b) and c), that for all values of \(\beta\), the steady-state histograms have sharp peaks at both \(N_{\textsf{  perf}}=0\) and \(N_{\textsf{  perf}}=1\); with one biased at high \(\beta\) and one biased at low. At one mode, the fraction of perfect genomes is close to zero, and the rate at which new perfect genomes appear in the process is also low. This slow rate leads to the fraction of perfect genomes increasing only slowly, even if perfect genomes are fit and have relatively long lifetimes. At the other mode, most of the genomes in the population are perfect; the marginal frequency of 1 is high at all loci, and new perfect genomes are bred with high probability; occasional imperfect genomes are quickly removed from the population. Both of these modes may be metastable. Transitions from the high-fitness mode to low-fitness and the reverse occur only through large fluctuations from a relatively stable mode. 

We explore this metastability by asking how many iterations of the birth/death process are necessary for the system to move from one peak to the other, thus exploring the entire sequence space. While our system is too complex to compute these quantities, we can make the following crude approximation.

To pass between the two peaks in probability, the system must pass through a probability minimum \(N_{\textsf{  perf}}=N^{\textsf{  min}}_{\textsf{  perf}}\). Therefore, the amount of times the system passes between the two peaks must be constrained by the number of times the system passes through this probability minimum.

For a given number of birth-death cycles \(T_{\textsf{  tot}}\), \(T^{\textsf{  min}}_{\textsf{  tot}}=\Ps (N_{\textsf{  perf}}=N^{\textsf{  min}}_{\textsf{  perf}})T_{\textsf{  tot}}\) are spent in the probability minima \(N_{\textsf{  perf}}=N^{\textsf{  min}}_{\textsf{  perf}}\). We define the average amount of cycles (or lifetime) spent in the probability minima each time the system passes through it as \(T^{\textsf{  min}}\). Therefore the amount of times the system passes through the probability minima is \(\frac{T^{\textsf{  min}}_{\textsf{  tot}}}{T^{\textsf{  min}}}\).

 \(T_{\textsf{  esc\,bound}}=T_{\textsf{  tot}}\frac{T^{\textsf{  min}}}{T^{\textsf{  min}}_{\textsf{  tot}}}=\frac{T^{\textsf{  min}}}{\Ps (N_{\textsf{  perf}}=N^{\textsf{  min}}_{\textsf{  perf}})}\) is then the average number of birth-death cycles between each instance of moving through the probability minima. It is a lower bound on the time it takes for the system to move from one peak to the other, what we term the ``escape time" \(T_{\textsf{  esc}}>T_{\textsf{  esc\, bound}}=\frac{T^{\textsf{  min}}}{\Ps (N_{\textsf{  perf}}=N_{\textsf{  perf}}^{\textsf{  min}}})\). 

Given the difficulty of estimating \(T^{\textsf{  min}}\), we make the following simplification. We count every birth/death cycle after which the system is in the probability minima separately, even if they are consecutive, setting \(T^{\textsf{  min}}=1\). This simplification overestimates the number of times the system passes through the probability minima and therefore underestimates the average escape time bound \(T_{\textsf{  esc\, bound}}\). However, since \(T_{\textsf{  esc\, bound}}\) is a lower bound, this makes the bound weaker but does not negate it. Our bound is now therefore \(T_{\textsf{  esc}}>\frac{1}{\Ps (N_{\textsf{  perf}}=N^{\textsf{  min}}_{\textsf{  perf}})}\). 

In figure \ref{PerfectFitnessHistograms}a) we plot this bound in red. We see that as \(\beta\) increases and the average probability of a genome being perfect increases, the escape time from the metastable state and, therefore, the time to reach the stationary state diverges.

The blue curve in figure \ref{PerfectFitnessHistograms}a) shows the expected fraction of perfect genomes as a function of $\beta$. At the value of $\beta$ where the expected fraction of perfects is approx $\frac{1}{2}$, the population spends approximately equal expected time in both modal regimes, with rare switches between them. The variance of $\phi$ for this population is maximized: $\phi(\pop)$ is either close to 0 or close to $N$. This section of the blue curve also has the highest gradient, according to the fluctuation theorem of section \ref{sec:fitnessfluctuations}, equation \ref{eq:fluctuationtheorem}.

\subsection{Perfect, prefix, and sum fitness: mean $\phi$}

\begin{figure}[h]
    \centering
    \includegraphics[scale=0.5]{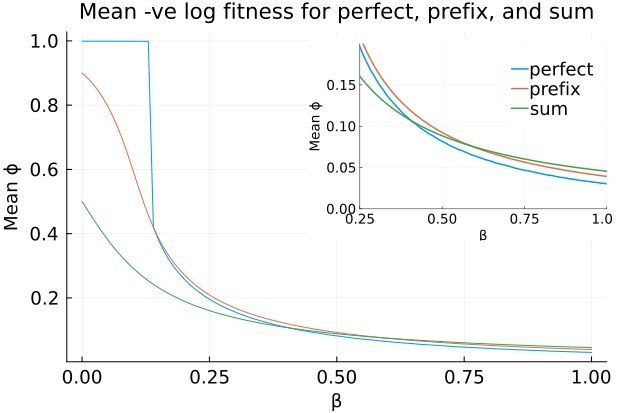}
    \caption{Mean $\phi$ plotted against $\beta$ for three fitness functions in the infinite population limit. Mutation rate $u=0.01$. Lower values of $\phi$ correspond to higher fitness. For low $\beta$, Perfect fitness selection gives the highest expected $\phi$, but for high $\beta$, perfect fitness selection induces the lowest mean $\phi$; indeed, the ordering of the three fitness functions reverses with the more intense selection at high $\beta$. An intuitive explanation is that in Sum and Prefix fitness, there are many configurations with near-optimal fitness, so the prior total $\Pu$ of these configurations increases mean $\phi$. }
    \label{fig:meanvslog}
\end{figure}

The Perfect, Prefix, and Sum fitness functions then have the property that for all genomes $\bg$, $\fperfect(\bg) \le \fprefix(\bg) \le \fsum(\bg)$. One might rashly surmise that the mean fitnesses of genomes in the stationary distributions for the three fitness functions would be in the same order. In fact, the behavior of mean fitness (or, as we plot, mean $\phi$, the negative log fitness weight) is more complicated. 

Figure \ref{fig:meanvslog} shows the mean $\phi$ at stationarity in the infinite population limit, plotted against $\beta$, the selection intensity. Mean $\phi$  is plotted for evolution with Perfect, Prefix, and Sum fitness.   For low values of $\beta$, perfect fitness selection is not powerful enough to maintain any significant fraction of perfect genomes in the population. However, at a critical value of $\beta$, mean $\phi$ changes discontinuously.   For larger values of $\beta$, perfect fitness selection maintains the lowest mean $\phi$ (highest mean fitness), followed by prefix selection, with sum selection giving the lowest mean fitness of the three. The inset graph shows no simple relationship between expected mean fitness, fitness function, and $\beta$.

An intuitive explanation of this effect is that Sum fitness allows many configurations of the population in which most genomes have slightly sub-optimal fitness. Here, even at high $\beta$, much of the prior probability mass of $\Pu$ is on these sub-optimal configurations, reducing the mean fitness weight of the population. In contrast, with $\fperfect$, imperfect genomes are unfit, and if $\beta$ is sufficiently high, nearly all genomes are perfect at stationarity. 

The effect of the structure of the fitness function on the mean fitness is of biological interest. Our models can perform exact calculations of these biologically relevant quantities for an endless number of fitness functions with relative ease. Further work would be to examine additional fitness functions.

\begin{figure}[h]
    \centering
    \includegraphics[scale=0.4]{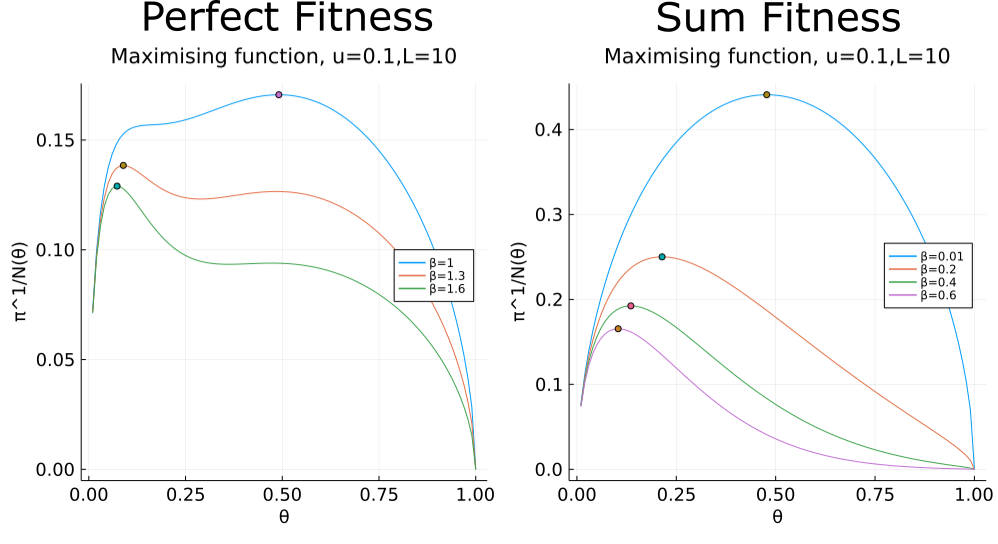}
    \caption{Shows $\maximiser(\theta,\fperfect)$ and $\maximiser(\theta,\fsum)$ plotted against $\theta$ for different values of $\beta$. In the $\delta$-function method of determining the stationary distribution in the infinite population limit, the optimal $\theta^*$ is the maximizer of $\maximiser$. The left-hand graph shows that $\maximiser( \theta,\fperfect)$ has two maxima. These change relative value as $\beta$ increases: at a critical value of $\beta$, the global maximizer jumps from one maximum to the other so that there is a discontinuous change in $\theta^*$. The right-hand graph shows that there is only one maximum of $\maximiser( \theta, \fsum)$ for all values of $\beta$ so that $\theta^*$ is a continuous function of $\beta$. }
    \label{fig:maximfitness}
\end{figure}

\subsection{Error correcting codes can improve population fitness}
\label{sec:errorcorrecting}

It has been appreciated for fifty years that many mutations are neutral in their effects on fitness. The consequences of neutrality were first addressed theoretically in Kimura's \emph{neutral theory} of evolution \cite{Kimura:1968:Evolutionary}. The question of whether evolution can come to shape the probability that mutations are neutral has been the subject of more recent theory on the evolution of \emph{mutational robustness} \cite{Nimwegen:Crutchfield:and:Huynen:1999, Bornberg-Bauer:and:Chan:1999:Modeling}. The collection of genotypes with equal fitness that are mutationally connected form a \emph{neutral network}. The topological structure of the neutral network determines where the population will evolve, and typically the population evolves to where a greater proportion of mutations are neutral.

Here we generate neutral networks and examine the consequences for evolution by proposing that the organism has a means of ``error correction'' in how the genotype maps to the phenotype, such that single mutations away from a focal genotype are compensated to give the same fitness. We employ classical Hamming codes to generate this neutral network.

Hamming codes, developed in 1950 by \cite{hamming1950error}, are error-correcting codes that detect and correct a single corrupted bit in a binary string. A Hamming codeword --- a binary sequence of length $2^r-1$, for some integer $r$, contains $r$ `check bits' used in decoding, and $2^r - r - 1$ `message bits', which contain the message to be sent. The simplest Hamming code consists of codewords with three bits, each containing one message bit and two check-bits. There are two codewords: 000 and 111. If a single bit of either of these codewords is corrupted, for example, 000 might be corrupted to 001, we can easily detect and correct this corruption. This correction is possible because the corrupted bit differs from two uncorrupted bits: the system takes a `majority vote' of the bits to correct the error. We refer to this as a 1/3 Hamming Code --- a code with one message bit and codewords of length 3.  

One bit of information in a genome can be encoded in two ways. Firstly, as a single allele, that may take the values 1 or 0. Alternatively, as three different alleles, each taking the values 1 or 0, where the encoded bit is the majority-vote value of the three alleles. For example, consider a pair of organisms. Organism A encodes one bit naively using a single allele, and organism B uses the majority vote of 3 alleles. The fitness functions for organism B depend only on the decoded bit: organism B is fit if the 3 alleles decode to 1 and unfit if they decode to 0. Does organism B have any advantage over organism A in expected fitness under identical selective conditions? 

There are Hamming codes with 4 message bits and 7 bit codewords, as well as 11/15 and 26/31. Given that a Hamming code can correct exactly one error, in our calculations and simulations, we use the ``One Error" fitness function for our encoded genomes (length \(2^{r}-1\) ) and the ``Perfect" fitness function for our shorter compact genomes (length \(2^{r}-r-1\)).

\begin{figure}
    \centering
    \includegraphics[scale=0.3]{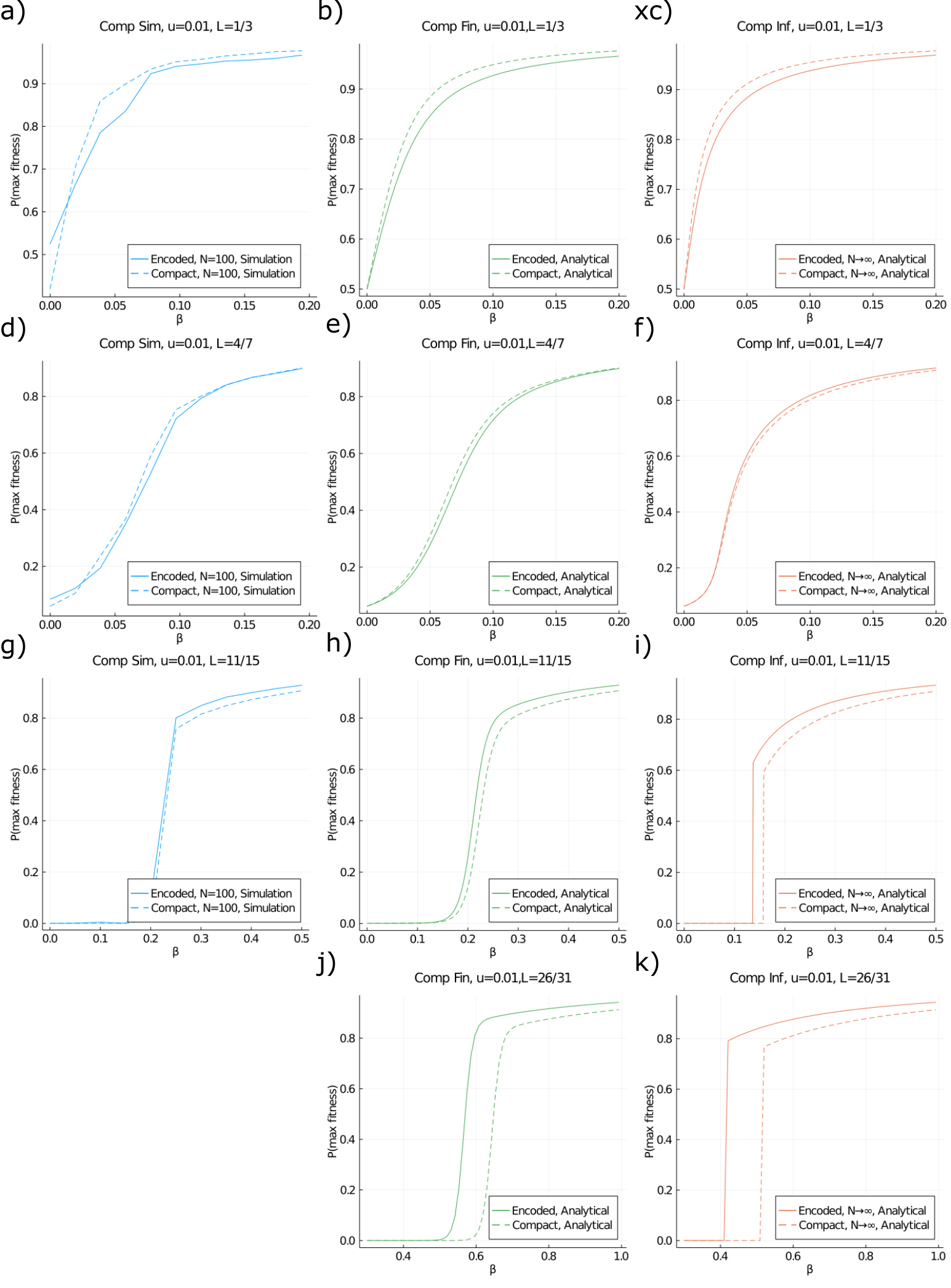}
    \caption{This shows the relative performance of Hamming decoding vs. compact representation for genomes of different lengths. The first column shows the results of simulations; the second column shows theoretical values for a finite population $N=100$, and the third column shows theoretical values for an infinite population. In all graphs, we fix \(u=0.01\) and plot the fraction of perfect genomes against \(\beta\). The first row compares a compact genome with one bit against the smallest possible Hamming code that decodes 3 bits to 1: compact representations are better for this code length under all conditions we tried. Row 2 compares a compact coding of 4 bits against a Hamming code of 7 bits that decodes to 4 bits: the performance of the two encodings is similar. Rows 3 and 4 show code lengths of 11 bits vs. the 15 bit Hamming code and a code length of 26 bits vs. a 31 bit Hamming code. The Hamming code shows an increasing advantage for these longer code lengths, giving a higher fraction of perfect decoded words for values of $\beta$ above a threshold.}
    \label{HammingGraphs}
\end{figure}

In figure \ref{HammingGraphs}a-c, we plot the properties of the compact \(L=1\) and encoded \(L=3\) case. We directly compare two systems, one with the ``perfect" fitness function at the compact length and one with the ``one error" fitness function at the encoded length. An encoded genome with one or fewer errors will decode onto a perfect genome. We plot the results of simulations for the two systems with \(N=100\) in (a), the analytical finite population result for the two systems with \(N=100\) in (b),  and the analytical solution for the two systems with \(N\rightarrow\infty\) in (c). We do this for compact \(L=4\) encoded \(L=7\) (d-f), compact \(L=11\) encoded \(L=15\) (g-i) and compact \(L=26\) encoded \(L=31\) (j-k). We omit the simulations for the last case because the system takes too long to settle to the stationary state. Using the method outlined in section \ref{EstimateEscape}, we can calculate the minimum number of birth-death cycles required for the system with compact \(L=26\) encoded \(l=31\) to move between the various stable states of the system once, which is essential to finding the true stationary state. For the compact message, at the transition at \(u=0.01\) \(\beta=0.55\), the system would take \(>1.61\times10^9\) birth/death iterations to move between the various stationary states. For the encoded message at the transition at \(\beta=0.62\), the system would take \(>4.14\times 10^{10}\) to move between the various stationary states. Note that the escape time at the transition will be among the shortest; increasing or decreasing \(\beta\) would make the escape time longer.

In fig. \ref{HammingGraphs}a-c or e-f, we plot the results for systems with short messages of length one or four. Here there are a comparable number of error-correcting bits to message bits; two error-correcting bits for a message of length one and three error-correcting bits for a message of length four. Tripling or almost doubling the message length to correct a single error bit does not help transmit the message more accurately because the longer genomes are subject to more mutations. In fact, for the shortest messages \(L=1\), it is better to send the message compactly than to encode it, and for messages of length \(L=4\), the encoded and compact cases lie nearly on top of each other.

However, for longer messages of length 11 or 26 (fig. \ref{HammingGraphs} g-i or j-k), there are much fewer error correcting bits relative to length, around 35\% and 20\% of the original message length, respectively. Thus correcting an error for a much smaller relative message length is a significant improvement. We observe the encoded population performing visibly better in regions of moderate up to high \(\beta\) in both the finite and infinite number of genomes cases. There is a significant difference between the compact and encoded cases at the transition point.

Previous work\cite{watkins2008selective} has suggested that sexual reproduction may perform better transmitting genomes in which the information is spread over a long genome. This analysis of Hamming codes shows that, in principle, an organism endowed with the ability to decode its binary genome with a Hamming code could have higher expected mean fitness at stationarity.  

We do not suggest that any organism actually uses a Hamming code to produce mutational robustness. The Hamming code investigated here generates a particular neutral network, which we show reduces the genetic load caused by mutation. These results suggest that evolutionary dynamics may produce neutral networks that utilize cooperative interactions between multiple alleles in ways reminiscent of error-correcting codes. 

\section{Discussion} 
We have introduced a simple continuous-time evolutionary model in which the fitness of a genome corresponds to its expected lifetime in the population.  

This model abstracts some aspects of biology and simplifies others. For example, to model sexual reproduction, we have abstracted the principle that each new genome consists of copies of genetic material fairly sampled from the current population --- but for mathematical simplicity, copying is from the entire population instead of from two parents only. Does this matter? For population-genetic questions, this simplification goes too far. However, we can use the mathematical simplicity we gain with the clean factorization of the stationary distribution to answer other questions that are hard or impossible to address otherwise. 

Our model leaves out more biological details than others, but the mathematical simplicity allows us to address new questions. As with any modelling abstraction, the question to ask is whether the mechanisms that we have left out, such as breeding from only two parents, give evolution essential computational power, or whether these omitted mechanisms are accidents of biology of little computational significance. We leave this for further work. 

\medskip

This research was supported by grant number (FQXi Grant number FQXi-RFP-IPW-1913) from the Foundational Questions Institute and Fetzer Franklin Fund, a donor advised fund of Silicon Valley Community Foundation.

\bibliographystyle{unsrt}
\bibliography{Bibliography.bib}

\beginsupplement

\section{Supplementary Material}

\section{Simulations}

To show the success of our numerical solutions, we compare them to a simulation of the exchangeable Moran process. In this section and all sections containing graphs, we convert from the mathematically elegant quantity \(\alpha\) to the more intuitive quantity of the mutation rate \(u\). In the setting where \(\alpha_{0}=\alpha_{1}=\alpha\) then \(\alpha_{i}=n\frac{u}{1-u}\). We set up a population with \(n=30\) and \(l=10\). We run separate runs for set values of \(u\) and \(\beta\). Recall that \(u\) is the probability that any given site in any given reproductive event will be a mutation rather than being drawn from the exchangeable parent distribution: the larger \(u\), the greater the amount of noise in the population. Equally \(\beta\) quantifies the relative cost of divergence from a local fitness maximum. Small \(\beta\) populations have more opportunity to explore the whole parameter space to find the global fitness maxima but are less stable in any fitness maxima it reaches. Large \(\beta\) populations are more stable in a local fitness maxima but have less opportunity to explore the whole range of genome space to find the global fitness maxima.

We seed the population with genomes where every locus has a 50:50 chance of being correct or incorrect. Then, we run the population for between \(5000-40000\) generations depending on how quickly the population converges to the stationary state, with each generation having \(n\) selection and replacement cycles per locus. Finally, we average the probability of a given genome decoding to perfect over the final 10\% of iterations. We do this process for each of the fitness functions described above.

\begin{figure}
    \includegraphics[scale=0.5]{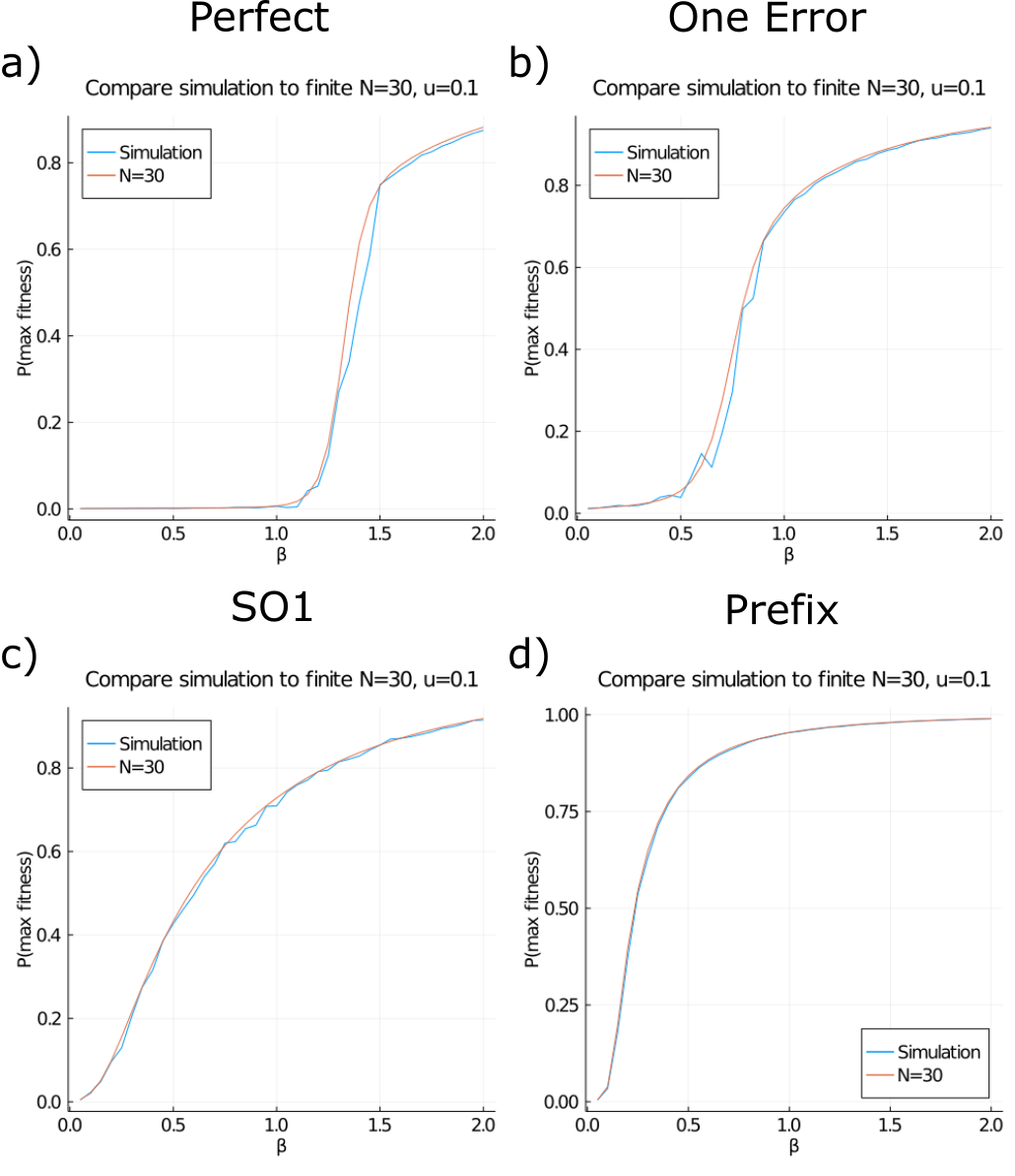}
    \caption{We compare simulation and numerical results for systems with \(n=30\) and \(l=10\). For the simulations, we seed the population with genomes where every locus has a 50:50 chance of being correct or incorrect. We run the population for between \(5000-40000\) generations depending on how quickly the population converges to the stationary state, with each generation having \(n\) selection and replacement cycles per locus. We average the probability of a given genome decoding to perfect over the final 10\% of iterations. We do this process for the Sum, Perfect, Prefix and One Error fitness functions. We also solve these models directly using the recursion methods described in section \ref{FF}. These are compared directly in this figure. We observe that in general, as \(\alpha\) increases for fixed beta, the probability of a given genome having the maximum possible fitness decrease, with a very sharp change in the perfect fitness case. As \(\beta\) increases at fixed \(\alpha\), the probability of a genome having the maximum possible fitness increases, with a sharp change in all three fitness cases. We observe excellent agreement in general, with some deviation during the sharp probability changes due to the simulated system having insufficient time to find the long-term stationary state.}
    \label{finitesim}
\end{figure}

In figure \ref{finitesim}, we compare the simulations with the numerical results. We can see excellent agreement for low values of \(u\), with some noise at higher \(\alpha\), meaning the population struggles to converge on the stationary state even at long run times. For \(l=10\), the number of genomes with maximum fitness corresponds to around \(1\%\) and \(0.1\%\) of states for the one error and perfect fitness function, respectively. As such, the one error and perfect fitness simulations may fail to match the numerical result where there are sharp changes in the probability of a given genome having max fitness even for relatively long run times.

\end{document}